\documentstyle[12pt,epsfig,amsfonts]{article}
\begin{document}

\voffset -0.4 true cm
\hoffset 1.3 true cm
\topmargin 0.0in
\evensidemargin 0.0in
\oddsidemargin 0.0in
\textheight 8.7in
\textwidth 6.9in
\parskip 10 pt

\newcommand{\be}{\begin{equation}}
\newcommand{\ee}{\end{equation}}
\newcommand{\bea}{\begin{eqnarray}}
\newcommand{\eea}{\end{eqnarray}}
\newcommand{\beas}{\begin{eqnarray*}}
\newcommand{\eeas}{\end{eqnarray*}}

\font\cmss=cmss10
\def\half{{1 \over 2}}
\def\identity{{\rlap{\cmss 1} \hskip 1.6pt \hbox{\cmss 1}}}
\def\laplace{{\kern1pt\vbox{\hrule height 1.2pt\hbox{\vrule width 1.2pt\hskip
  3pt\vbox{\vskip 6pt}\hskip 3pt\vrule width 0.6pt}\hrule height 0.6pt}
  \kern1pt}}
\def\scriptlap{{\kern1pt\vbox{\hrule height 0.8pt\hbox{\vrule width 0.8pt
  \hskip2pt\vbox{\vskip 4pt}\hskip 2pt\vrule width 0.4pt}\hrule height 0.4pt}
  \kern1pt}}
\def\slash#1{{\rlap{$#1$} \thinspace /}}
\def\roughly#1{\raise.3ex\hbox{$#1$\kern-.75em\lower1ex\hbox{$\sim$}}}
\def\complex{{\hbox{\cmss C} \llap{\vrule height 7.0pt
  width 0.4pt depth -.4pt \hskip 0.5 pt \phantom .}}}
\def\real{{\hbox{\cmss R} \llap{\vrule height 6.9pt width 0.4pt
  depth -.1pt \hskip 0.6 pt \phantom .}}}
\def\integer{{\rlap{\cmss Z} \hskip 1.8pt \hbox{\cmss Z}}}

\font\cmsss=cmss8
\def\C{{\hbox{\cmsss C}}}
\def\bigC{{\hbox{\cmss C}}}
\def\gs{{g^2_{YM}}}
\def\gssb{{g^2_{YM} \sqrt{\beta}}}
\def\gsb{{g^2_{YM} \beta}}
\def\gfb{{g^4_{YM} \beta}}

\begin{titlepage}
\begin{flushright}
{\small IASSNS-HEP-99/51} \\
{\small CU-TP-953} \\
{\small PUPT-1855} \\
{\small hep-th/9910001}
\end{flushright}

\begin{center}

\vspace{1mm}

{\Large \bf Approximations for strongly-coupled supersymmetric quantum mechanics}

\vspace{2mm}

Daniel Kabat${}^{1,2}$ \ and Gilad Lifschytz${}^3$

\vspace{1mm}

${}^1${\small \sl School of Natural Sciences} \\
{\small \sl Institute for Advanced Study, Princeton, NJ 08540} \\
{\small \tt kabat@sns.ias.edu}

\vspace{1mm}

${}^2${\small \sl Department of Physics} \\
{\small \sl Columbia University, New York, NY 10027}

\vspace{1mm}

${}^3${\small \sl Department of Physics} \\
{\small \sl Princeton University, Princeton, NJ 08544} \\
{\small \tt gilad@viper.princeton.edu}

\end{center}

\vskip 0.3 cm

\noindent
We advocate a set of approximations for studying the finite
temperature behavior of strongly-coupled theories in 0+1 dimensions.
The approximation consists of expanding about a Gaussian action, with
the width of the Gaussian determined by a set of gap equations.  The
approximation can be applied to supersymmetric systems, provided that
the gap equations are formulated in superspace.  It can be applied to
large-$N$ theories, by keeping just the planar contribution to the gap
equations.

\noindent
We analyze several models of scalar supersymmetric quantum mechanics,
and show that the Gaussian approximation correctly distinguishes
between a moduli space, mass gap, and supersymmetry breaking at strong
coupling.  Then we apply the approximation to a bosonic large-$N$
gauge theory, and argue that a Gross-Witten transition separates the
weak-coupling and strong-coupling regimes.  A similar transition
should occur in a generic large-$N$ gauge theory, in particular in
0-brane quantum mechanics.

\end{titlepage}

\section{Introduction}

Recent developments have made it clear that at a non-perturbative
level string or M-theory often has a dual description in terms of
large-N gauge theory \cite{BFSS,ads}.  This has many interesting
implications, both for string theory and for the behavior of large-N
gauge theory.  These developments have been reviewed in
\cite{reviews}.

A general property of these dualities is that semiclassical gravity
can be used only in regimes where the gauge theory is strongly
coupled.  For certain terms in the effective action, which are
protected by supersymmetry \cite{SethiSternPaban}, a weak coupling
calculation in the gauge theory can be extrapolated to strong coupling
and compared to supergravity \cite{bb,bbpt,OkawaYoneya,WatiMark}.  But
in general non-perturbative methods must be used to analyze the gauge
theory in the supergravity regime.  Although this reflects the power
of duality, as providing a solution to strongly coupled gauge theory,
it is also a source of frustration.  For example, one would like to
use the duality to understand black hole dynamics in terms of gauge
theory
\cite{BFKS,KlebanovSusskind,Halyo,HorowitzMartinec,Li,DMRR,BFKS2,TsLi,BFK,OhtaZhou,LiMartinec,Lowe,KL1,KL2}.
But progress along these lines has been hampered by our inadequate
understanding of gauge theory.

Thus we would like to have direct control over the gauge theory at
strong coupling.  Such control is presumably easiest to achieve in a
low-dimensional setting.  The specific system we have in mind is
0-brane quantum mechanics \cite{ClaudsonHalpern}, in the temperature
regime where it is dual to a 10-dimensional non-extremal black hole
\cite{imsy}.  The dual supergravity makes definite predictions for the
partition function of this gauge theory.  In the temperature regime
under consideration, the partition function is in principle given by
the sum of planar diagrams.  Although directly summing all planar
diagrams seems like a hopeless task, one might hope for a reasonable
approximation scheme, which can be used to resum a sufficiently large
class of diagrams to see agreement with supergravity.

In this paper we develop a set of approximations which, we hope, can
eventually be used to do this.  Essentially we self-consistently resum
an infinite subset of the perturbation series.  These methods have a
long history in many-body physics \cite{ManyBody} and field
theory \cite{DolanJackiw,CJT}, and have been used to study QCD
\cite{QCD,QCD2}, large-N gauge theory \cite{EngelhardtLevit}, and even
supersymmetry breaking \cite{Zanon}.  The basic idea is to approximate
the theory of interest with a Gaussian action.  The Gaussian action is
(in general) non-local in time, with a variance that is determined by
solving a set of gap equations.\footnote{We refer to gap equations
even when studying models without a mass gap.}  One can systematically
compute corrections to this approximation, in an expansion about the
Gaussian action.  To study large-N theories in this formalism, one
keeps only the planar contribution to the gap equations.  To study
supersymmetric systems, we will see that one needs to formulate the
gap equations in superspace: loop corrections to the auxiliary fields
must be taken into account in order for the approximation to respect
supersymmetry.

An outline of this paper is as follows.  In section 2 we illustrate
the approximation in the simple context of $0+0$ dimensional $\phi^4$
theory, and discuss various ways of formulating gap equations.  We
also discuss (but do not resolve) the difficulties with implementing
gauge symmetry in the Gaussian approximation.  In section 3 we turn to
a series of quantum mechanics problems with ${\cal N} = 2$
supersymmetry but no gauge symmetry, and show that the Gaussian
approximation captures the correct qualitative strong-coupling
behavior present in these simple systems.  In section 4 we discuss
large-$N$ gauge theories in $0+1$ dimensions, and argue that
generically a Gross-Witten transition is present, which separates the
weak-coupling and strong-coupling phases.  In section 5 we apply the
Gaussian approximation to two large-$N$ quantum mechanics problems: a
bosonic gauge theory, and a supersymmetric matrix model with a moduli
space.  We discuss the relevance of these results for 0-brane quantum
mechanics.

\section{Formulating the Gaussian approximation}

\subsection{A simple example: $\phi^4$ theory in $0+0$ dimensions}
Our approach to studying strongly-coupled low-dimensional systems can
be illustrated in the following simple context.  Consider $0+0$ dimensional
$\phi^4$ theory, with action
\be
\label{phi4:action}
S = {1 \over g^2}\left(\half \phi^2 + {1 \over 4} \phi^4\right)\,.
\ee
The partition function is given in terms of a Bessel function, by
\be
\label{phi4:Z}
Z = e^{-BF} = \int_{-\infty}^\infty d\phi \, e^{-S} = 
{1 \over \sqrt{2}} e^{1/8g^2} K_{1/4}\left({1 \over 8 g^2}\right)\,.
\ee
Expanding this result for weak coupling one obtains an asymptotic
series for the free energy.
\be
\label{phi4:weak}
\beta F = - \half \log \left(2 \pi g^2\right) + {3 \over 4} g^2 - 3 g^4
+ {99 \over 4} g^6 + {\cal O}(g^8)
\ee
This series can of course be reproduced by doing conventional
perturbation theory in the coupling $g^2$.

But suppose we are interested in the behavior of the free energy at
strong coupling.  From the exact expression we know that this is given
by
\be
\label{phi4:strong}
\beta F = - \half \log g - \log {\pi \over \Gamma(3/4)} + {\cal O}(1/g)\,.
\ee
Is there some approximation scheme that reproduces this result?
Clearly working to any finite order in conventional perturbation
theory is hopeless.  But it turns out that by resumming a subset of
the perturbation series one can do much better.  To this end consider
approximating the theory (\ref{phi4:action}) with the following Gaussian
theory.
\bea
Z_0 & = & e^{-\beta F_0} = \int_{-\infty}^\infty d\phi \, e^{-S_0} \nonumber \\
S_0 & = & {1 \over 2 \sigma^2} \phi^2
\label{phi4:gaussian}
\eea
The width of the Gaussian $\sigma^2$ is for the moment left arbitrary.  By writing
\[
Z = \int d\phi \, e^{-S_0} e^{-(S - S_0)}
\]
one obtains the identity
\be
\label{phi4:exact}
\beta F = \beta F_0 \,\, - <e^{-(S - S_0)} - 1>_{\C,0}
\ee
where the subscript $\bigC,0$ denotes a connected expectation value in
the Gaussian theory (\ref{phi4:gaussian}).  By expanding this
identity in powers of $S - S_0$ we obtain a reorganized perturbation
series\footnote{There is a real advantage to keeping at least the
first two terms in the reorganized perturbation series: one is free to add
a constant term to $S_0$, which may depend on the couplings or
temperature but is independent of the fields.  Any such additive
ambiguity cancels between $BF_0$ and $<S - S_0>_0$.}
\be
\label{phi4:series}
\beta F = \beta F_0 + <S - S_0>_0 - \half <(S - S_0)^2>_{\C,0} + \cdots\,.
\ee
The first few terms are given by
\beas
& & \beta F_0 = -\half \log 2 \pi\sigma^2 \\
& & <S - S_0> = \half \left({\sigma^2 \over g^2} - 1 \right) + {3 \sigma^4 \over 4 g^2} \\
& & - \half <\left(S - S_0\right)^2>_{\C,0} = - {1 \over 4} \left( {\sigma^2 \over g^2} - 1\right)^2
- {3 \sigma^4 \over 2 g^2} \left({\sigma^2 \over g^2} - 1\right) - {3 \sigma^8 \over g^4} \,.
\eeas

The identity (\ref{phi4:exact}) holds for any choice of $S_0$.  But we wish to
truncate the series (\ref{phi4:series}), and the truncated series
depends on the value of $\sigma^2$.  So we need a prescription for
fixing $\sigma^2$.  One reasonable prescription is as follows.
Consider the Schwinger-Dyson equation\footnote{This equation follows
from noting that an infinitesimal change of variables $\phi
\rightarrow \phi + \delta \phi$ leaves the integral $\int_{-
\infty}^\infty d\phi \, \phi \, e^{-S(\phi)}$ invariant.}
\[
<\phi {\delta S \over \delta \phi}> = < {1 \over g^2} \left(\phi^2 + \phi^4\right)> = 1 \,.
\]
This is an exact relation among correlation functions in the full
theory.  It seems reasonable to ask that the same relation hold in the
Gaussian theory, that is, to require that $<{1 \over g^2}
\left(\phi^2 + \phi^4\right)>_0 = 1$.  This implies that $\sigma^2$ obeys
the gap equation
\be
\label{phi4:gap}
{1 \over \sigma^2} = {1 \over g^2} + {3 \sigma^2 \over g^2}\,.
\ee
This equation has a simple diagrammatic interpretation, shown in
Fig. 1: it self-consistently resums certain self-energy corrections to
the $\phi$ propagator.

\begin{figure}
\epsfig{file=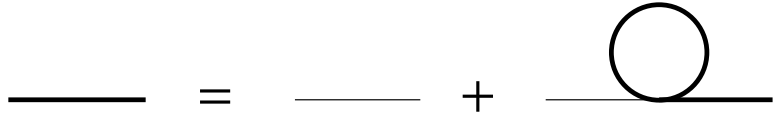}
\caption{The gap equation for the $\phi$ propagator.  Heavy lines are
the dressed propagator $\sigma^2$ and thin lines are the bare
propagator $g^2$.}
\end{figure}

At weak coupling the gap equation implies $\sigma^2 = g^2 + {\cal
O}(g^4)$ and thus the leading term in the Gaussian approximation is
\beas
\beta F_0 & = & - \half \log 2 \pi g^2 + {\cal O}(g^2)\,.
\eeas
This matches the weak coupling behavior seen in (\ref{phi4:weak}).
At weak coupling the higher order terms in the reorganized
perturbation series are small, suppressed by increasing powers of the
coupling.

Of course it is not surprising that the Gaussian approximation works
well at weak coupling, since at weak coupling the action (\ref{phi4:action})
is approximately Gaussian.  But what is rather remarkable is that the
Gaussian approximation also gives good results at strong coupling,
where the $\phi^4$ term in $S$ dominates.  At strong
coupling the gap equation gives $\sigma^2 = {g \over \sqrt{3}} + {\cal
O}(1)$ and thus the leading term in the Gaussian approximation is
\[
\beta F_0 = - \half \log g - \half \log {2 \pi \over \sqrt{3}} + {\cal O}(1/g)\,.
\]
Thus $\beta F_0$ has the same leading strong-coupling behavior as the
exact result (\ref{phi4:strong}).  Moreover, by reorganizing the
perturbation series we have {\em tamed the strong-coupling behavior of
perturbation theory}.  That is, all the higher order terms appearing
in the expansion (\ref{phi4:series}) are only ${\cal O}(1)$ at strong
coupling.
\beas
& & <S - S_0> = - {1 \over 4} + {\cal O}(1/g) \\
& & - \half <(S-S_0)^2>_{\C,0} = - {1 \over 12} + {\cal O}(1/g) \\
& & \qquad \qquad \qquad \vdots
\eeas
Indeed the first few terms of the reorganized perturbation series fall
off nicely with the order.  This provides an intrinsic reason to
expect that truncating the reorganized perturbation series at low
orders should provide a good approximation to the exact
result.\footnote{Although we haven't investigated the issue, it seems
too much to hope that the reorganized perturbation series is actually
convergent.}

\subsection{Varieties of gap equations}

Consider a set of degrees of freedom $\phi_i$, governed by an action
$S(\phi_i)$.  We wish to approximate this system with a simpler action
$S_0(\phi_i)$.  For the most part we will take $S_0$ to be Gaussian in
the fundamental fields.
\[
S_0 = \sum_i {1 \over 2 \sigma_i^2} \phi_i^2
\]
But in some situations, in particular for gauge theories, we will be
led to use a more general action, which is exactly soluble but not strictly
Gaussian, of the form
\[
S_0 = \sum_a {1 \over \lambda_a} {\cal O}_a\,.
\]
Here the $\lambda_a$ are adjustable parameters, and the ${\cal O}_a$ are
composite operators built out of the fundamental fields.

In either case, the adjustable parameters appearing in $S_0$ will be
chosen to satisfy a set of gap equations.  There are several
prescriptions for writing down gap equations.  All the gap equations
given below have one feature in common: they correspond to critical
points of an effective action.\footnote{For gap equations which follow
from a variational principle, the gap equations give a global minimum
of the effective action.  But in general the gap equations only give a
critical point.}  We now present several prescriptions for writing
down gap equations.

\begin{itemize}

\item{Lowest-order Schwinger-Dyson gap equations}

For a generic degree of freedom $\phi_i$ one has the Schwinger-Dyson
equation
\be
\label{exactSD}
<\phi_i {\delta S \over \delta \phi_i}> = 1
\ee
(no sum on $i$).  This is an exact relation among correlation
functions in the full theory, which can be derived by the standard
procedure sketched in section 2.1.  It seems natural to choose the
parameters of $S_0$ in such a way that the same relation holds between
correlation functions evaluated in $S_0$.  That is, a natural set of gap
equations corresponds to demanding that for each $i$
\be
\label{lowestSD}
<\phi_i {\delta S \over \delta \phi_i}>_0 = 1\,.
\ee

This gap equation can be obtained from an effective action, as
follows.  Assuming that $S_0$ is strictly Gaussian, and writing $S$ as
a polynomial in the fundamental fields, one can show that the gap
equations (\ref{lowestSD}) are equivalent to requiring
\be
\label{lowestSD:Leff}
{\partial \over \partial \sigma_i^2} \left(\beta F_0 + <S - S_0>_0\right) = 0\,.
\ee
That is, these gap equations make the first two terms of the
perturbation series for $\beta F$ stationary with respect to
variations of the $\sigma_i^2$.

\item{Operator gap equations}

It is natural to ask that expectation values of the operators
${\cal O}_a$ agree when calculated in $S_0$ and in $S$:
\be
\label{OperatorCondition}
<{\cal O}_a> = <{\cal O}_a>_0.
\ee
For example, when discussing gauge theories, we will introduce the
Wilson loop operator ${\cal O} = {\rm Tr} P e^{i \oint A}$.  Then the
condition (\ref{OperatorCondition}) demands that Wilson loops agree
when computed in $S_0$ and in $S$.  As another example, suppose that
$S_0$ is Gaussian, with ${\cal O}_i = \phi_i^2$.  Then
(\ref{OperatorCondition}) means choosing the Gaussian widths
$\sigma_i^2$ to be the exact 2-point functions of the full theory.

As it stands (\ref{OperatorCondition}) is not a useful equation.  But
it can be rewritten as
\be
\label{OperatorCondition2}
<{\cal O}_a e^{-(S - S_0)}>_{\C,0} = <{\cal O}_a>_0
\ee
and expanded in powers of $S - S_0$.  At leading order this condition
is trivially satisfied.  But at first order we get the gap
equation\footnote{It is tempting to try to improve this gap equation
by expanding (\ref{OperatorCondition2}) to higher orders in $S - S_0$.
But this rarely seems to lead to useful gap equations.}
\be
\label{OperatorLeadingOrder}
<{\cal O}_a (S - S_0)>_{\C,0} = 0\,.
\ee

To relate this to an effective action, note the identities
\beas
& & {\partial \over \partial \lambda_a} \beta F_0 = - {1 \over \lambda_a^2} <{\cal O}_a>_0 \\
& & {\partial \over \partial \lambda_a} <S - S_0>_0 = {1 \over \lambda_a^2} <{\cal O}_a (S - S_0)>_{\C,0}
+ {1 \over \lambda_a^2} <{\cal O}_a>_0\,.
\eeas
Thus the gap equation (\ref{OperatorLeadingOrder}) is equivalent to requiring
\be
\label{Operator:Leff}
{\partial \over \partial \lambda_a} \left(\beta F_0 + <S - S_0>_0\right) = 0\,.
\ee
Note that if $S_0$ is strictly Gaussian then these operator gap
equations reduce to (\ref{lowestSD:Leff}).

\item{Variational gap equations}

In some cases gap equations can be obtained from a variational principle.  For
purely bosonic theories one has a bound on the free energy, that
\[
\beta F \leq \beta F_0 + <S-S_0>_0\,.
\]
Thus one can formulate a variational principle, based on minimizing
the right hand side of the inequality with respect to $\lambda_a$
\cite{Feynman}.  Note that the resulting gap equations are equivalent
to (\ref{Operator:Leff}).

\end{itemize}

The gap equations that we have discussed so far are all essentially
equivalent: they follow from varying $\beta F_0 + <S - S_0>_0$.
Unfortunately, they will not be sufficient to analyze many of the
theories of interest in this paper.  For example, suppose that $S_0$
is purely Gaussian, and that $S$ only has non-trivial 3-point
couplings.  Such couplings do not contribute to $\beta F_0 + <S -
S_0>_0$, so the gap equations discussed above are not sensitive to the
interactions.  But then, if some field is massless at tree level,
$\beta F_0$ will suffer from an infrared divergence.

To resolve these difficulties, we turn to the

\begin{itemize}
\item{One-loop gap equations}

The gap equation of section 2.1 resummed an infinite set of
Feynman diagrams, involving one-loop self-energy corrections to the
propagator.  This can be extended to theories with both 3-point
and 4-point interactions, as follows.  Write $S = S_{II} + S_{III} +
S_{IV} + \cdots$ as a sum of 2-point, 3-point, 4-point, $\ldots$
couplings, and assume that $S_0$ is strictly Gaussian.  Consider the quantity
\[
I_{\rm eff} = \beta F_0 + <S_{II} - S_0>_0 + <S_{IV}>_0 - \half <(S_{III})^2>_{\C,0}\,.
\]
Requiring ${\partial \over \partial \sigma_i^2} I_{\rm eff} = 0$
gives a set of gap equations which resum all one-loop self-energy
corrections to the propagators, built from both 3-point and 4-point
vertices.  These gap equations are illustrated in Fig.~2.  They are
sufficient to cure the infrared divergences for most theories of
interest.  The quantity $I_{\rm eff}$ can be identified with the 2PI effective action of Cornwall, Jackiw and Tomboulis \cite{CJT},
as calculated in a two-loop approximation.

\end{itemize}

\begin{figure}
\epsfig{file=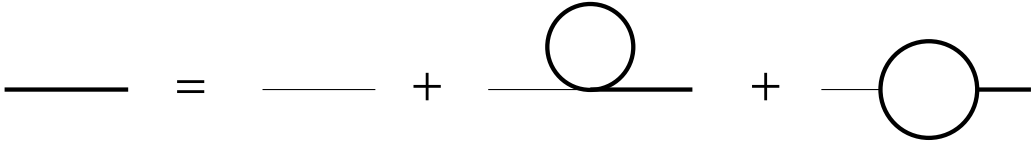}
\caption{One-loop gap equations.  Heavy lines are dressed
propagators and thin lines are bare propagators.}
\end{figure}

In general it does not seem possible to give a prescription for
choosing the best set of gap equations.  Indeed it is not even clear
what `best' means -- it may depend on what one wishes to calculate.
For example, in some cases, the variational principle gives the
optimal gap equations to use when estimating the free energy.  A
different set of gap equations will give a worse estimate of the free
energy, but may give a better estimate of other quantities, such as
2-point correlations.  The only general advice we can offer is that
the gap equations must be chosen to cure all the infrared divergences
which are present in the model.  If possible, the gap equations should
also respect all the symmetries of the model.\footnote{This becomes
problematic when dealing with gauge theory, as we discuss in the next
subsection.}

\subsection{Difficulties with gauge symmetry}

There is a serious difficulty which one encounters when applying the
Gaussian approximation to a gauge theory.  The underlying gauge
invariance implies Ward identities, which relate different Green's
functions.  But the Gaussian approximation often violates these
identities, and this can lead to inconsistencies.

To illustrate the problem, consider an action $S(\phi_i)$ which
only has 3-point and 4-point couplings -- for example ordinary
Yang-Mills theory.  We assume that the one-point functions $<\phi_i>$
vanish.  Then the exact Schwinger-Dyson equations $<\phi_i {\delta S
\over \delta \phi_i}> = 1$ can be rewritten in terms of proper (1PI)
vertices as shown in Fig.~3 \cite{QCD,QCD2,BjorkenDrell}.

\begin{figure}
\epsfig{file=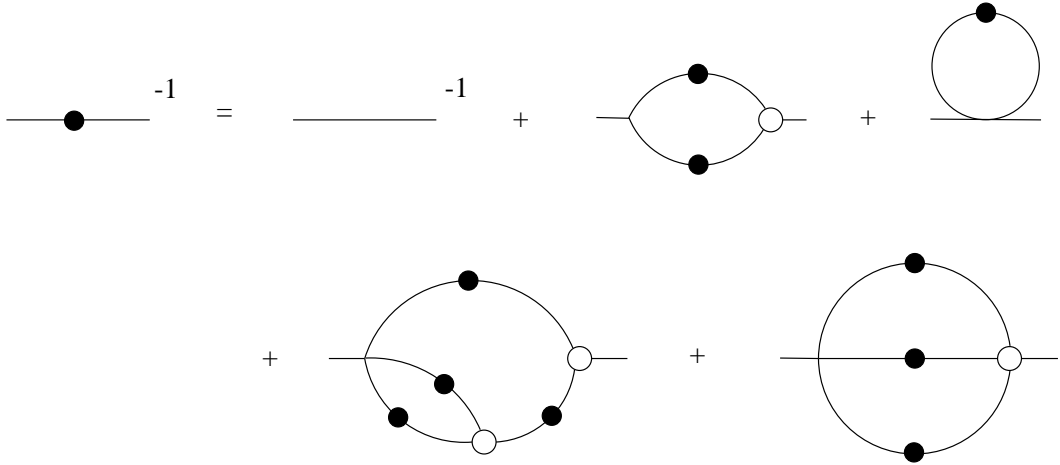}
\caption{Schwinger-Dyson equations for a theory with 3-point and
4-point couplings but no tadpoles.  The solid blobs are dressed
propagators; the empty circles are 1PI vertices.  All external lines
are amputated.}
\end{figure}

The one-loop gap equations which we discussed in the previous
subsection are a truncation of this system.  They correspond to
dropping the two-loop contributions to the Schwinger-Dyson equations
and approximating the dressed 1PI 3-point vertex with a bare vertex.
This truncation works well for many theories, as we will see.  But in
an abelian gauge theory, for example, a Ward identity relates the
dressed 3-point vertex to a dressed propagator.  This makes it
inconsistent to work with dressed propagators but bare vertices; the
inconsistency shows up in the fact that the one-loop gap equations
with bare vertices do not have a solution.\footnote{This result was
obtained in conjunction with David Lowe.}  Methods for dealing with
this difficulty have been proposed \cite{QCD,QCD2}, but they do not
seem to be compatible with manifest supersymmetry.  An adequate
analysis of the 0-brane quantum mechanics calls for a more
satisfactory treatment of this issue \cite{KLL}.  In section 5, when
we apply the Gaussian approximation to a bosonic large-$N$ gauge
theory, we will dodge the issue of Ward identities, by avoiding the
use of the full set of one-loop gap equations.

\section{Scalar models with supersymmetry}

In this section we apply the Gaussian approximation to study the
strong-coupling behavior of some quantum mechanics problems having
${\cal N} = 2$ supersymmetry but no gauge symmetry.  The goal is
twofold: to learn how to apply the Gaussian approximation to a
supersymmetric system, and to test the Gaussian approximation against
some of the well understood strong-coupling dynamics present in these
simple systems.  The main lesson is that the Gaussian approximation
works well, provided that it is formulated in a way which respects
supersymmetry.  Effectively this requires formulating the Gaussian
approximation in superspace: gap equations must be introduced for the
auxiliary fields.

\subsection{A model with a moduli space}

We begin by considering a model with three scalar superfields and a
superpotential $W = \Phi_1 \Phi_2 \Phi_3$.  The Minkowski space action
is (see Appendix A for supersymmetry conventions)
\[
S_M = {1 \over g^2} \int dt \, \half \dot{\phi}_a \dot{\phi}_a
+ i \bar{\psi}_a \dot{\psi}_a + \half f_a f_a - \half f_a \phi_b \phi_c s_{abc}
+ \phi_a \bar{\psi}_b \psi_c s_{abc}
\]
where $s_{abc} = +1$ if $a,b,c$ are distinct, $0$ otherwise.
Non-perturbatively it is known that the model has a moduli space of
vacua\footnote{in the sense of the Born-Oppenheimer approximation}
localized along the $\phi_1$, $\phi_2$, $\phi_3$ coordinate axes.
Thus the spectrum of the Hamiltonian is continuous from zero, and the
free energy falls off like a power law at low temperatures.

In studying this model, our goal is to show that the Gaussian
approximation captures this power-law behavior of the free energy at
low temperatures.  This is not a trivial result, because going to low
temperatures is equivalent to going to strong coupling: the coupling
constant $g^2$ is dimensionful, $g^2 \sim ({\rm length})^{-3}$, so the
dimensionless effective coupling is $g_{\rm eff}^2 = g^2 / T^3$.  We
henceforth adopt units in which $g^2 = 1$.

We use an imaginary time formalism, Wick rotating according to
\[
S_E = -i S_M \qquad \tau = i t \qquad f_{a\,E} = -i f_{a\,M} \,.
\]
Note that we have to Wick rotate the auxiliary fields in order to make
the quadratic terms in the action well-behaved in Euclidean space.
Expanding the fields in Fourier modes
\beas
\phi_a(\tau) & = & {1 \over \sqrt{\beta}} \sum_{l \in \integer}
\phi^a_l \, e^{-i 2 \pi l \tau / \beta} \\
\psi_a(\tau) & = & {1 \over \sqrt{\beta}} \sum_{r \in \integer
+ {1 \over 2}} \psi^a_r \, e^{-i 2 \pi r \tau / \beta} \\
f_a(\tau) & = & {1 \over \sqrt{\beta}} \sum_{l \in \integer}
f^a_l \, e^{- i 2 \pi l \tau / \beta}
\eeas
the Euclidean action is
\beas
S_E & = & \half \sum_l \left({2 \pi l \over \beta}\right)^2 \phi_l^a \phi_{-l}^a
+ \sum_r -i {2 \pi r \over \beta} \bar{\psi}_r^a \psi_r^a + \half \sum_l f_l^a f_{-l}^a \\
& & \qquad + {i \over 2 \sqrt{\beta}} \sum_{l+m+n = 0} \phi_l^a \phi_m^b f_n^c s_{abc}
- {1 \over \sqrt{\beta}} \sum_{l-r+s = 0} \phi_l^a \bar{\psi}_r^b \psi_s^c s_{abc}\,.
\eeas
For a Gaussian action we take (the $R$-parity symmetry
discussed in Appendix A forbids any $\phi$ -- $f$ mixing)
\[
S_0 =  \sum_l {1 \over 2 \sigma_l^2} \phi_l^a \phi_{-l}^a
     + \sum_l {1 \over 2 \tau_l^2} f_l^a f_{-l}^a
     + \sum_r {1 \over g_r} \bar{\psi}_r^a \psi_r^a \,.
\]
Next we need to choose a set of gap equations.  This model only has cubic
interactions, so a reasonable choice is the one-loop gap equations discussed in
section 2.2.  These read
\bea
{1 \over \sigma^2_l} & = & \left({2 \pi l \over \beta}\right)^2
+ {2 \over \beta} \sum_{m+n = l} \sigma_m^2 \tau_n^2
- {2 \over \beta} \sum_{r-s = l} h_r h_s \nonumber \\
\label{moduli:gap}
{1 \over \tau_l^2} & = & 1 + {1 \over \beta} \sum_{m+n = l} \sigma_m^2 \sigma_n^2 \\
{1 \over h_r} & = & {2 \pi r \over \beta} + {2 \over \beta} \sum_{l+s = r} \sigma_l^2 h_s
\nonumber
\eea
where $h_r = -i g_r$ is real.  Note that time-reversal invariance
requires the symmetry properties
\[
\sigma^2_{-l} = \sigma^2_l \qquad \tau^2_{-l} = \tau^2_l \qquad h_{-r} = - h_r \,.
\]
Meanwhile the first few terms in the reorganized perturbation series for the
free energy are
\beas
&& \beta F_0 = - {3 \over 2} \sum_l \log \sigma_l^2 - {3 \over 2} \sum_l \log \tau_l^2
+ {3 \over 2} \sum_r \log h_r^2 \\
&& <S - S_0>_0 = {3 \over 2} \sum_l \left(({2 \pi l \over \beta})^2 \sigma_l^2 - 1\right)
+ {3 \over 2} \sum_l \left(\tau_l^2 - 1\right) - 3 \sum_r \left( {2 \pi r \over \beta} h_r - 1 \right) \\
&& - \half <(S - S_0)^2>_{\C,0} = - {3 \over 4} \sum_l \left(({2 \pi l \over \beta})^2 \sigma_l^2 - 1 \right)^2
- {3 \over 4} \sum_l \left(\tau_l^2 - 1\right)^2 \\
&& \qquad \qquad + {3 \over 2} \sum_r \left({2 \pi r \over \beta} h_r - 1\right)^2
+ {3 \over 2 \beta} \sum_{l + m + n = 0} \sigma_l^2 \sigma_m^2 \tau_n^2
- {3 \over \beta} \sum_{l - r + s = 0} \sigma_l^2 h_r h_s\,.
\eeas

\begin{figure}
\epsfig{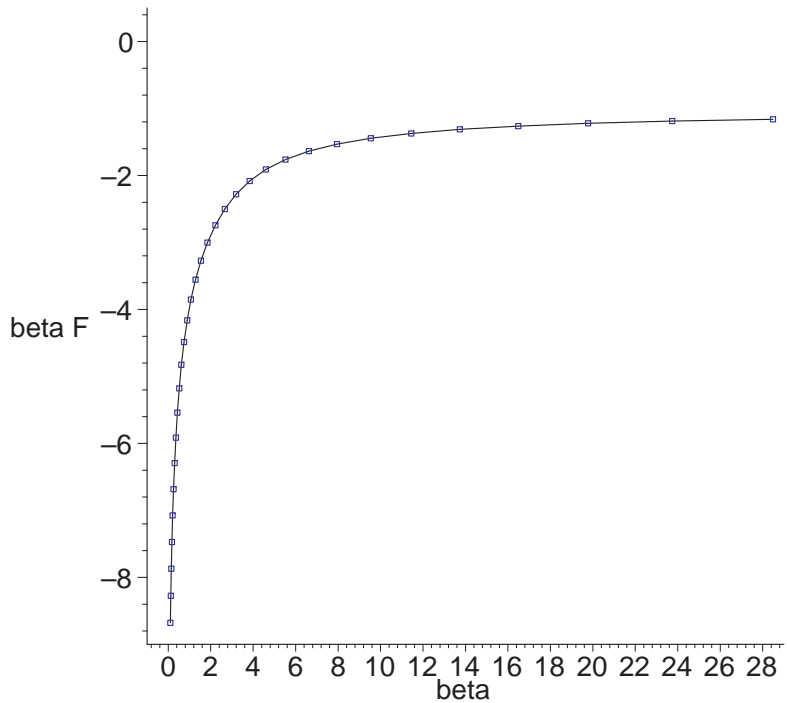}
\caption{$\beta F$ vs.~$\beta$ for the $\Phi_1 \Phi_2 \Phi_3$ model,
as given by the sum of the first three terms of the reorganized
perturbation series.  Numerical calculations were performed at the
indicated points.}
\end{figure}

\begin{figure}
\epsfig{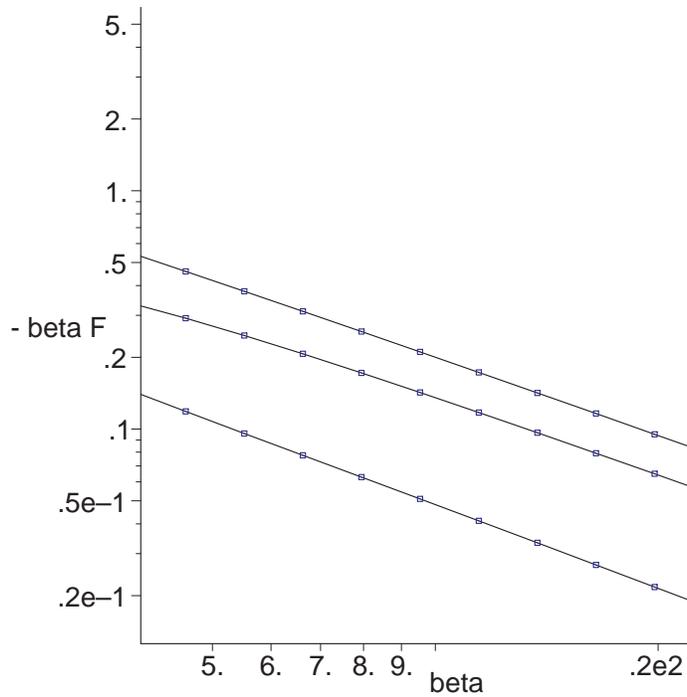}
\caption{Plots of $(-)$ individual terms in the free energy
vs.~$\beta$ in the low temperature regime.  The scale is
$\log$--$\log$, and we shifted $\beta F_0$ by a constant to display
its power-law behavior.  The top curve is $-<S - S_0>_0$, the middle
curve is $+\half<(S - S_0)^2>_{\C,0}$, and the bottom curve is
$-(\beta F_0 + 1.04)$.}
\end{figure}

To solve the gap equations (\ref{moduli:gap}) one has to resort to
numerical methods.  An effective technique is to start by solving the
gap equations analytically at high temperatures, where the loop sums
are dominated by the bosonic zero modes $\sigma_{l=0}^2$,
$\tau^2_{l=0}$.  One can then repeatedly use the Newton-Raphson method
\cite{NumericalRecipes} to solve the system at a sequence of
successively smaller temperatures, using the solution at inverse
temperature $\beta$ as the starting point for Newton-Raphson at $\beta
+ \Delta \beta$.  Further details on the numerical algorithm are given
in appendix B.

We have solved this system numerically up to $\beta = 28.5$,
corresponding to a dimensionless coupling $g_{\rm eff}^2 = g^2 \beta^3
\approx 10^4$.  The resulting free energy
\be
\label{moduli:first3}
\beta F \approx \beta F_0 + <S - S_0>_0 - \half <(S - S_0)^2>_{\C,0}
\ee
is shown in Fig.~4.  From the plot it seems clear that the energy of
the system -- given by the slope of $\beta F$ -- vanishes as the
temperature goes to zero, so the Gaussian approximation has captured
the fact that supersymmetry is unbroken in this model.  Note that loop
corrections to the auxiliary field propagators were crucial in
obtaining this result -- if we had formulated a Gaussian approximation
just for the physical degrees of freedom, the approximation itself
would have explicitly broken supersymmetry.

Moreover, in the low-temperature regime, the numerical results for
the individual terms in the free energy are well fit by
\bea
& & \beta F_0 = -1.04 - 0.705 \beta^{-1.16} \nonumber \\
\label{moduli:numerics}
& & <S - S_0>_0 = -2.58 \beta^{-1.11} \\
& & - \half <(S - S_0)^2>_{\C,0} = -1.77 \beta^{-1.11} \,. \nonumber
\eea
This behavior can be seen in Fig.~5, where we have plotted these
quantities on a $\log$--$\log$ scale.

Evidently $\beta F_0$ approaches a non-zero constant at low
temperatures, indicating an effective ground-state degeneracy of the
system.  It would be interesting to understand whether such a
degeneracy is really present, or whether it is perhaps an artifact of
the Gaussian approximation.\footnote{The spectrum of the Hamiltonian
is continuous from zero, so the usual argument that $\beta F
\rightarrow 0$ as $\beta \rightarrow \infty$ does not apply.  In fact
it is not even clear whether the zero-temperature partition function for
this model is well-defined, since it may depend on how one regulates
divergences coming from the infinite volume of moduli space.}

What is more interesting for our purposes is that the subleading
behavior of $\beta F_0$, as well as the corrections $<S - S_0>_0$ and
$- \half <(S - S_0)^2>_{\C,0}$, all vanish like a power law at low
temperatures, with about the same exponent.  It seems plausible
that all the higher-order corrections will also follow roughly the same
power law.  If this is the case then the Gaussian approximation
provides a good estimate of the exponent.  The coefficient in front of
the power law is more difficult to obtain, but one might hope that
keeping just the first three terms as in (\ref{moduli:first3}) is not
such a bad approximation.  This leads to a numerical fit
\[
\beta F \approx -1.04 - 4.77 \beta^{-1.09} \,.
\]

\subsection{A model with a mass gap}

Next we consider a model containing a single scalar superfield with
superpotential $W = {1 \over 4} \Phi^4$, corresponding to the Minkowski action
\[
S_M = {1 \over g^2} \int dt \, \half \dot{\phi}^2 + i \bar{\psi} \dot{\psi} + \half f^2
- f \phi^3 + 3 \phi^2 \bar{\psi} \psi\,.
\]
Non-perturbatively this
theory is known to have a unique supersymmetric vacuum localized near
$\phi = 0$ \cite{Witten}.  In this section we study the finite
temperature behavior of this model using the Gaussian approximation,
and show that it reproduces this known behavior in the limit of zero
temperature.  Moreover we will use the Gaussian approximation to
obtain an estimate for the energy gap to the first excited state of
the system.  This is non-trivial because, just as in the previous
section, going to low temperature is equivalent to going to strong
coupling: the dimensionless effective coupling is $g_{\rm eff}^2 =
g^2/T^2$.  From now on we set $g^2 = 1$.

The Euclidean action in Fourier modes reads
\beas
S_E & = & \sum_l \half \left({2 \pi l \over \beta}\right)^2 \phi_l \phi_{-l}
+ \sum_r -i {2 \pi r \over \beta} \bar{\psi}_r \psi_r + \sum_l \half f_l f_{-l} \\
& & \quad + {i \over \beta} \sum_{l+m+n+o = 0} \phi_l \phi_m \phi_n f_o - {3 \over \beta}
\sum_{l+m-r+s = 0} \phi_l \phi_m \bar{\psi_r} \psi_s\,.
\eeas
For a Gaussian action we take
\[
S_0 = \half \sum_l \left(\matrix{\phi_l & f_l \cr} \right)
\left(\matrix{\sigma^2_l & \rho_l \cr \rho_l & \tau^2_l \cr}\right)^{-1}
\left(\matrix{\phi_{-l} \cr f_{-l} \cr} \right)
+ \sum_r {1 \over g_r} \bar{\psi}_r \psi_r\,.
\]
Note that one has to allow the fields $\phi$ and $f$ to mix in the
Gaussian action, because the superpotential breaks the $R$-parity symmetry
discussed in appendix A.  Thus we have the correlators
\beas
<\phi_l \phi_m>_0 & = & \sigma_l^2 \delta_{l+m} \qquad <\phi_l f_m>_0 = \rho_l \delta_{l+m} \\
\noalign{\vskip 0.2 cm}
<f_l f_m>_0 & = & \tau_l^2 \delta_{l+m} \qquad <\bar{\psi_r} \psi_s>_0 = - g_r \delta_{rs} \,.
\eeas

For this model the one-loop gap equations are identical to the lowest-order
Schwinger-Dyson gap equations.  In either case one obtains the system
\bea
\left(\matrix{\sigma^2_l & \rho_l \cr \rho_l & \tau^2_l \cr}\right)^{-1} & = &
\left(\matrix{(2\pi l/\beta)^2 & 0 \cr 0 & 1 \cr}\right) +
\left(\matrix{{6 i \over \beta} \sum_m \rho_m + {6 \over \beta} \sum_r g_r &
{3 i \over \beta} \sum_m \sigma_m^2 \cr 
{3 i \over \beta} \sum_m \sigma_m^2 & 0 \cr}\right) \nonumber \\
\noalign{\vskip 0.2 cm}
{1 \over g_r} & = & -i {2 \pi r \over \beta} - {3 \over \beta} \sum_m \sigma_m^2
\label{gap:eqns}
\eea
The solution to the gap equations has the form
\beas
\sigma^2_l & = & {1 \over (2 \pi \l / \beta)^2 + m_b^2} \\
\noalign{\vskip 0.1 cm}
\rho_l & = & - i m_f \sigma_l^2 \\
\noalign{\vskip 0.1 cm}
\tau_l^2 & = & 1 - m_f^2 \sigma_l^2 \\
\noalign{\vskip 0.1 cm}
g_r & = & {1 \over -i 2 \pi r / \beta - m_f}
\eeas
parameterized by effective bose and fermi masses $m_b$, $m_f$.  It is convenient to
introduce the size of the state
\bea
R_{\rm rms}^2 & = & <\phi(\tau)^2> \nonumber \\
\label{mass:Rrms2}
& = & {1 \over \beta} \sum_l \sigma_l^2 \\
& = & {1 \over 2 m_b \tanh(\beta m_b/2)} \,. \nonumber
\eea
The gap equations then imply consistency conditions which fix $m_b$, $m_f$
as functions of the temperature.
\bea
m_b^2 & = & 27 R_{\rm rms}^4 - 3 \tanh(3 \beta R_{\rm rms}^2/2) \nonumber \\
\label{mass:consistancy}
m_f & = & 3 R_{\rm rms}^2
\eea
Meanwhile the first two terms in the expression for the free energy are
\beas
& & \beta F_0 = - \half \sum_l \log \left(\sigma_l^2 \tau_l^2 - \rho_l^2\right)
+ \half \sum_r \log \vert g_r \vert^2 \\
& & <S - S_0>_0 = \half \sum_l \left(({2 \pi l \over \beta})^2 \sigma_l^2 - 1\right)
+ \half \sum_l (\tau_l^2 - 1) + \sum_r \left(i {2 \pi r \over \beta} g_r + 1\right) \\
& & \qquad\qquad\qquad + {3 i \over \beta} \sum_l \sigma_l^2 \sum_m \rho_m
+ {3 \over \beta} \sum_l \sigma_l^2 \sum_r g_r\,.
\eeas
With the help of the gap equations this reduces to
\bea
\beta F_0 \,\, + <S - S_0>_0 & = & \log(2 \sinh(\beta m_b/2))
- \log(2 \cosh(\beta m_f /2)) \nonumber \\
\label{mass:betaF}
& & - \half \beta m_b^2 R_{\rm rms}^2 + {9 \over 2} \beta R_{\rm rms}^6 \,.
\eea

The nice thing is that if you expand for low temperatures
(equivalently strong coupling -- recall that the dimensionless
coupling is $g_{\rm eff}^2 = g^2/T^2$) then the solution to the
consistency conditions (\ref{mass:Rrms2}), (\ref{mass:consistancy}) is
\[
\beta m_b = \sqrt{3 \over 2} \, \beta + \sqrt{6} \, \beta e^{-\beta \sqrt{3/2}}
+ {\cal O}\left(e^{-\beta \sqrt{6}}\right)\,.
\]
The corresponding free energy (\ref{mass:betaF}) is
\be
\label{mass:LowTemp}
\beta F_0 \,\,+ <S - S_0>_0 = -2 e^{-\beta \sqrt{3/2}}
+ {\cal O}\left(e^{-\beta \sqrt{6}}\right) \,.
\ee
But this is exactly the low temperature expansion one would expect for the
free energy of a system with a unique zero energy ground state (a
`short multiplet' of the ${\cal N} = 2$ supersymmetry) plus a
degenerate pair of first excited states with energy $\sqrt{3/2}$ (a
`long multiplet' of ${\cal N} = 2$).

We conclude by examining the second-order corrections to the free energy of this model.
After imposing the gap equations one finds that
\[
- \half <(S-S_0)^2>_{\C,0} = \frac{1}{9}\beta m_f^3
+ \frac{9}{\beta^2}\sum_{l+m-r+s=0} \!\! \sigma^2_l \sigma^2_m g_r g_s
- \frac{12m_f^3}{\beta^2} \sum_{l+m+n+o=0} \!\! \sigma^2_l \sigma^2_m
\sigma^2_n \sigma^2_o \,.
\]
The three-loop sums are best evaluated in position space, using
($0 \leq \tau \leq \beta$)
\beas
{1 \over \beta} \sum_l \frac{1}{(2 \pi l/\beta)^2 +m_b^2} e^{-i2\pi l \tau/\beta}
& = & {\cosh m_b (\tau - \beta/2) \over 2 m_b \sinh (\beta m_b / 2)} \\
{1 \over \beta} \sum_{r}\frac{1}{- i 2\pi r / \beta - m_f} e^{-i2\pi r \tau/\beta}
& = & {e^{m_f(\tau - \beta/2)} \over 2 \cosh (\beta m_f / 2)},
\eeas
to obtain
\beas
& & - \half <(S-S_0)^2>_{\C,0} = \frac{1}{9}\beta m_{f}^{3}
- \frac{9(\beta m_{b}+\sinh \beta m_{b})}{16m_{b}^{3}
(1+\cosh\beta m_{f})\sinh^{2} \frac{\beta m_{b}}{2}} \\
& & \qquad \qquad \qquad -\frac{3\beta m_{f}^2}{4m_{b}^{5}
\sinh^{4}\frac{\beta m_{b}}{2}}(\frac{3}{8}\beta m_b +\frac{1}{2}
\sinh \beta m_{b}+\frac{1}{16}\sinh 2\beta m_{b}) \,.
\eeas
At low temperatures this reduces to
\begin{equation}
\label{mass:correction}
- \half <(S-S_{0})^2>_{\C,0} = - 2 \sqrt{{2 \over 3}} \, \beta e^{-\beta\sqrt{3/2}}\,.
\end{equation}
Although exponentially suppressed, this larger than the contribution of the
first two terms (\ref{mass:LowTemp}) by a factor of $\beta$.
Combining the results (\ref{mass:LowTemp}), (\ref{mass:correction}) we
see that the expression for the free energy at low temperatures has
the form
\[
\beta F = -2 e^{-\beta \sqrt{3/2}} \left(1 + \beta \sqrt{2/3} + \cdots\right)
+ {\cal O}\left(e^{-\beta \sqrt{6}}\right) \,.
\]
Presumably the second-order term can be interpreted as correcting the
energy gap to the first excited state.  It would be interesting to
find a set of gap equations for which the second order term in the
free energy is subdominant at low temperatures.

\subsection{A model which breaks supersymmetry}

Finally, we consider a model with a single scalar superfield and a
superpotential $W = {1 \over 3} \Phi^3$, corresponding to the Minkowski action
\[
S_M = {1 \over g^2} \int dt \, \half \dot{\phi}^2 + \half f^2 + i \bar{\psi} \dot{\psi}
- f \phi^2 + 2 \phi \bar{\psi} \psi\,.
\]
The dimensionless coupling constant is $g_{\rm eff}^2 = g^2 / T^3$, and we
henceforth set $g^2 = 1$.

This model is known to break supersymmetry \cite{Witten}.  This is an
effect which cannot happen at any finite order in perturbation theory,
but we will see that it is captured by the Gaussian approximation: the
ground state energy comes out positive.  This provides an interesting
contrast to the usual analysis of supersymmetry breaking via
instantons \cite{SalomonsonVanHolten}, which can be applied to this
model once a sufficiently large mass perturbation is introduced.  A
related analysis of supersymmetry breaking in the $O(N)$ $\sigma$-model
was carried out by Zanon \cite{Zanon}.

The $O(2)$ R-symmetry prohibits any $\phi$ -- $f$ mixing, and implies
$<\phi> = 0$, but allows a tadpole for $f$.  The expectation value of
$f$ is the order parameter for supersymmetry breaking.  To take it
into account we shift $f \rightarrow a + \tilde{f}$, with $<\tilde{f}>
= 0$.  The Euclidean action is\footnote{We Wick rotate the
fluctuation $\tilde{f}$ but not the expectation value $a$.}
\begin{eqnarray}
S_E & = & \frac{1}{2}\sum_{l}\left(\frac{2\pi l}{\beta}\right)^{2}\phi_{l}\phi_{-l}+
\frac{1}{2}\sum_{l}\tilde{f}_{l}\tilde{f}_{-l} - \sum_{r}i \frac{2\pi r}{\beta}\bar{\psi}_{r} \psi_{r}\nonumber\\
& & - \half \beta a^2 - i a \sqrt{\beta} \tilde{f}_0 + a \sum_l \phi_l \phi_{-l} \nonumber \\
& & + {i \over \sqrt{\beta}}\sum_{l+m+n = 0} \tilde{f}_l \phi_m \phi_n
- {2 \over \sqrt{\beta}} \sum_{l-r+s = 0} \phi_l \bar{\psi}_r \psi_s\,.
\label{sbeuc}
\end{eqnarray}
For a Gaussian action we take
\begin{equation}
S_0 = \sum_l {1 \over 2 \sigma_l^2} \phi_{l} \phi_{-l}
+ \sum_{l} {1 \over 2 \tau_l^2} \tilde{f}_{l} \tilde{f}_{-l}
+\sum_{r}{1 \over g_r} \bar{\psi}_{r}\psi_{r}\,.
\end{equation}
We will use the one-loop gap equations, which read ($h_r = - i g_r$)
\bea
\frac{1}{\sigma_{l}^{2}}&=&\left(\frac{2\pi l}{\beta}\right)^2 + 2a +
\frac{4}{\beta} \sum_{m+n=l}\sigma_{m}^{2}\tau_n^2
-\frac{4}{\beta}\sum_{r-s=l}h_r h_s \nonumber\\
\label{break:gap}
\frac{1}{\tau_{l}^{2}} & = & 1 + \frac{2}{\beta}\sum_{m+n=l}\sigma^{2}_{m}\sigma^{2}_{n} \\
\frac{1}{h_r} & = & \frac{2\pi r}{\beta} + \frac{4}{\beta}\sum_{l+s=r}\sigma^{2}_{l}h_s\,.
\nonumber
\eea
The value of $a$ is fixed by requiring $<\tilde{f}> = 0$.
At the one loop level this gives
\be
\label{break:a}
a = \frac{1}{\beta} \sum_{l}\sigma^{2}_{l}\,.
\ee
Note that in this approximation the order parameter for supersymmetry
breaking is the spread of the ground state wavefunction!  We
approximate the free energy as $\beta F \approx \beta F_0 + <S -
S_0>_0$, with
\beas
& & \beta F_0 = - \half \sum_l \log \sigma_l^2 - \half \sum_l \log \tau_l^2
                + \half \sum_r \log h_r^2 \\
& & <S - S_0>_0 =  \half \sum_l \left(({2 \pi l \over \beta})^2 \sigma_l^2 - 1 \right)
+ \half \sum_l \left(\tau_l^2 - 1\right) - \sum_r \left({2 \pi r \over \beta} h_r - 1\right) \\
& & \qquad \qquad \qquad - \half \beta a^2 + a \sum_l \sigma_l^2\,.
\eeas

\begin{figure}
\epsfig{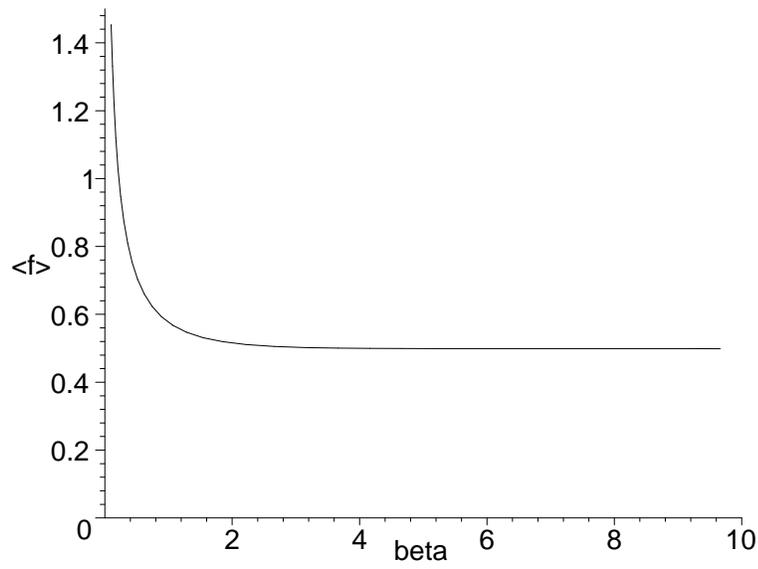}
\caption{Auxiliary field expectation value vs.~$\beta$ in the $\Phi^3$ model.}
\end{figure}

The gap equations for this model must be solved
numerically.  We show the result for $a = <f>$ in Fig.~6.  The key
point is that $<f>$ approaches a non-zero constant at low
temperatures, $<f> \approx 0.50$, so that supersymmetry is indeed
broken.  This is reflected in a positive slope of $\beta F$,
corresponding to positive vacuum energy.  The numerical results can be
fit at low temperatures by
\beas
& & \beta F_0 = -0.33 + 0.21 \beta \\
& & <S - S_0>_0 = -0.13 + 0.057 \beta \\
\noalign{\vskip 3 mm}
& & \beta F \approx -0.46 + 0.27 \beta \,.
\eeas
Note that, although the vacuum energy is non-zero, it is not given by
the classical formula $E = \half <f>^2$ -- we are at strong coupling and
quantum corrections are important.

To conclude, let us note that it is not really necessary to solve the full set of
one-loop gap equations to study this model.  Rather, it is a reasonable approximation
to truncate the gap equations (\ref{break:gap}) to
\beas
{1 \over \sigma_l^2} & = & \left({2 \pi l \over \beta}\right)^2 + 2a \\
{1 \over \tau_l^2} & = & 1 \\
{1 \over h_r} & = & {2 \pi r \over \beta}
\eeas
while keeping equation (\ref{break:a}) for $a$.  In this approximation the gap
equations are solved by
\[
a^3 = {1 \over 8 \tanh^2(\beta \sqrt{a/2})}
\]
while the free energy is given by
\[
\beta F \approx \beta F_0 + <S - S_0>_0 = \log \sinh (\beta \sqrt{a/2}) - \half \beta a^2\,.
\]
Note that at low temperatures $a \rightarrow 1/2$ and $\beta F
\rightarrow {3 \over 8} \beta$.

\section{Gauge dynamics and the Gross-Witten transition}

In this section we discuss a generic feature of large-$N$ gauge
theory: the existence of a Gross-Witten transition
\cite{GrossWitten}.\footnote{By Gross-Witten transition we mean a
large-$N$ phase transition triggered by a change in the eigenvalue
distribution, not necessarily a third order phase transition.}  We
will argue that such a transition separates the weak-coupling and
strong-coupling regimes of the gauge theory.

Suppose we have a $0+1$ dimensional $U(N)$ gauge theory at finite
temperature.  Although the gauge field $A_0(\tau)$ has no local
degrees of freedom, one can construct a non-local observable by
diagonalizing the Wilson line.
\[
U = P e^{i \int_0^\beta d\tau A_0} = \left(
\begin{array}{ccc}
e^{i \alpha_1} & & 0 \\
& \ddots & \\
0 & & e^{i \alpha_N}
\end{array}
\right)
\]
The eigenvalues $\alpha_1,\ldots,\alpha_N$ characterize the
gauge-invariant degrees of freedom contained in $A_0$ (up to
permutation by the Weyl group).

As discussed in \cite{Polhemus}, for most theories the eigenvalue distribution will be quite different
at high and low temperatures.  This can be understood as a competition
between entropy and energy.  The Haar measure for the eigenvalues,
corresponding to Wilson lines which are uniformly distributed over the
$U(N)$ group manifold, is given by \cite{Mehta}
\[
\prod_i d\alpha_i \prod_{i<j} \left(1 - \cos(\alpha_i - \alpha_j)\right)\,.
\]
The measure vanishes whenever two eigenvalues coincide.  So in the
absence of any other interactions, the eigenvalues tend to repel each
other, and spread out over the circle in order to maximize their
entropy.

On the other hand, energetic considerations often lead to attractive
interactions between eigenvalues.  As a simple example,
suppose we minimally couple the gauge field to a massive adjoint
scalar field.  Integrating out the scalar field produces an
additional contribution to the measure for the eigenvalues, given by
\bea
& & \prod_{i < j} \det{}^{-1} \left[-\left(\partial_\tau + {i \over \beta} (\alpha_i - \alpha_j)\right)^2
+ m^2\right] \nonumber \\
\label{FreePotential}
& & = \prod_{i < j} {1 \over 2\left(\cosh \beta m - \cos (\alpha_i - \alpha_j)\right)}\,.
\eea
In the high temperature regime $\beta m \ll 1$ this gives a strong
attraction between eigenvalues, and the eigenvalues will tend to
cluster to minimize their energy.  But at low temperatures the potential
(\ref{FreePotential}) flattens out, and entropy wins: the eigenvalues
will spread out around the circle.

This phenomenon seems to be quite generic: eigenvalues will tend to
cluster at high temperatures, and spread out at low
temperatures.  This should happen even in supersymmetric theories.\footnote{While
bosons generate an attractive potential which is minimized when
$\alpha_i - \alpha_j = 0$, fermions give rise to a repulsive potential
that repels away from $\alpha_i - \alpha_j = \pi$.}  Of course in
quantum mechanics at finite $N$ one has smooth crossover between these
two qualitatively different behaviors.  But in the large $N$ limit a
sharp Gross-Witten phase transition should occur when the eigenvalues
can first explore the entire circle.

We conclude with a few remarks on this phenomenon in the context of
D-brane physics; for a review of the phases of D-brane gauge theories
see \cite{MartinecReview}.
\begin{enumerate}
\item{}
The phenomenon is well-known, at least for D-branes which are wrapped
on a spatial circle.  It is responsible for the transition between
singly-wound and multiply-wound strings discussed in
\cite{DasMathur,MaldacenaSusskind}, and the related black hole
transitions discussed in \cite{SusskindGW}.  We've merely pointed out that
at finite temperature the same transition takes place in Euclidean
time: 0-brane worldlines can become multiply wound around the time
direction.  Only the multiply-wound phase corresponds to a black hole.
\item{}
In a T-dual picture the eigenvalues of the holonomy become the
positions of D-instantons on the dual circle.  At high temperatures
the dual circle is large, and the D-instantons tend to cluster.  But
at low temperatures, when the dual circle is small, the D-instantons
will delocalize.  Note that only in the low-temperature regime could
one hope to describe the configuration as a black hole, using a static
solution to the supergravity equations.

Let us make a crude estimate of the transition temperature from the
D-instanton point of view.  Consider a system of $N$ D-instantons in
IIB string theory with string coupling $g_B$ and string length
$l_s$.  Reasoning analogous to \cite{Susskind,Polchinski} shows that
in uncompactified space the D-instantons will cluster and fill out a
region of size $L \sim (g_B N)^{1/4} l_s$.  Compactifying the time
direction on a circle of size $\beta_B$, we expect the D-instantons to
delocalize if $\beta_B < L$.  Mapping to type IIA via
\[
\beta_A = \alpha' / \beta_B \qquad \qquad g^2_A = g^2_B \alpha' / \beta_B^2
\]
we conclude that the 0-brane worldlines are multiply-wound when
\be
\label{TransitionTemp}
\beta_A > \beta_{\rm crit} = (g_A N)^{-1/3} l_s \equiv (g^2_{YM} N)^{-1/3}\,.
\ee

\item{}
The transition at $\beta_{\rm crit}$ can be seen in several ways from
the supergravity point of view.  The background geometry of a system
of 0-branes in the decoupling limit is well-defined above
$\beta_{\rm crit}$.  Below $\beta_{\rm crit}$ the curvature at the
horizon exceeds the string scale, and the dual gauge theory becomes
weakly coupled \cite{imsy}.

The transition can also be identified with the Horowitz-Polchinski
correspondence point \cite{HorowitzPolchinski}.  Consider a
non-extremal black hole in ten dimensions with 0-brane charge.
Following \cite{HorowitzMartinec,DMRR}, one can map this to a
ten-dimensional Schwarzschild black hole by lifting to eleven
dimensions, boosting in the covering space, and recompactifying.  We
discuss this transformation in Appendix C.  If the original charged
black hole is at the critical temperature, then the equivalent neutral
black hole turns out to have a Schwarzschild radius close to the
string scale.  That is, the temperature $T_{\rm crit}$ maps to the
Horowitz-Polchinski correspondence point, at which a black hole is on
the verge of becoming an elementary string state.

\item{}
Although our primary interest is in quantum mechanics, it is natural
to suppose that similar transitions happen in other dimensions.  For
example, consider $3+1$ dimensional ${\cal N} = 4$ large-$N$
Yang-Mills on ${\Bbb R}^3 \times S^1$.  For any non-zero temperature
it seems natural to expect that a Gross-Witten transition occurs
when the 't Hooft coupling $\lambda = g^2_{YM} N$ is of order one.
\end{enumerate}

To conclude, we expect that Gross-Witten transitions are generically
present in large-$N$ gauge theories at finite temperature.  In
theories with a dual string interpretation this suggests that a phase
transition occurs, which separates the non-geometric phase from the
phase with a well-defined background geometry (see also
\cite{Vipul,VipulGilad}).

\section{Large-$N$ models in the Gaussian approximation}

In this section we use the Gaussian approximation to study some
large-$N$ matrix quantum mechanics problems.  One motivation for the
problems we consider is that they arise as subsectors of the 0-brane
quantum mechanics.  The 0-brane quantum mechanics can be written in
terms of ${\cal N} = 2$ superfields as \cite{KabatRey} (see Appendix
A)
\be
\label{D0:action}
S = {1 \over g^2_{\rm YM}} \int dt d^2\theta \, {\rm Tr} \Bigl\lbrace 
- {1 \over 4} \nabla^\alpha {\cal F}_i \nabla_\alpha {\cal F}_i
- \half \nabla^\alpha \Phi_a \nabla_\alpha \Phi_a
- {i \over 3} f_{abc} \Phi_a [\Phi_b,\Phi_c] \Bigr\rbrace\,.
\ee
Here ${\cal F}_i$ is the field strength constructed from a
gauge multiplet $\Gamma_\alpha$, the $\Phi_a$ are a collection of
seven adjoint scalar multiplets transforming in the ${\bf 7}$ of
$G_2$, and $f_{abc}$ is a suitably normalized\footnote{$f_{abc}
f_{abd} = {7 \over 2} \delta_{cd}$} totally antisymmetric
$G_2$-invariant tensor.\footnote{Under an $SO(2) \times SO(7)$
subgroup of the $SO(9)$ $R$-symmetry the physical fermions transform as an
${\bf 8}$ of $SO(7)$, which decomposes into ${\bf
7} \oplus {\bf 1}$ of $G_2$.  Thus one of the fermions becomes part of
the ${\cal N} = 2$ gauge multiplet, while the remaining seven transform
in the ${\bf 7}$ of $G_2$.}

In the next section we drop all the fermions, and (with nine scalar
fields) study the bosonic sector of this action.  In the following
section we drop the gauge multiplet $\Gamma_\alpha$, and study the
supersymmetric dynamics of the scalar multiplets $\Phi_a$.

\subsection{Bosonic large-$N$ gauge theory}

In this section we study a purely bosonic quantum mechanics problem:
the reduction of $d+1$ dimensional $U(N)$ Yang-Mills theory to $0+1$
dimensions.  We will study this theory at leading order for large $N$
using the Gaussian approximation.  The goal is to show that the
Gaussian approximation respects 't Hooft large-$N$ counting, and can
capture the Gross-Witten phase transition which we expect to be
present.  We will also learn a few useful lessons about 0-brane
quantum mechanics.

The degrees of freedom of this model consist of $d$ adjoint scalar
fields $X_i$, coupled to a gauge field $A_0$.  The Euclidean action is
\be
\label{bosonic:action}
S_E = {1 \over \gs} \int d\tau \, {\rm Tr} \left\lbrace \half D_\tau X^i D_\tau X^i - {1 \over 4}
[X^i,X^j] [X^i,X^j]\right\rbrace
\ee
where $D_\tau = \partial_\tau + i [A_0,\,\cdot\,]$.  We set
$\partial_\tau A_0 = 0$, so that $A_0 = {\rm const.} \equiv {1 \over
\sqrt{\beta}} A_{00}$.  Corresponding to this gauge choice we must
introduce a system of periodic ghost fields $\alpha$, $\bar{\alpha}$.
In Fourier modes the complete action is\footnote{Note that our gauge
condition leaves a residual global $U(N)$ gauge symmetry unbroken.  We
have suppressed the corresponding ghost zero mode $\alpha_{l = 0}$.}
\beas
S_E & = & {1 \over 2 \gs} \sum_l \left({2 \pi l \over \beta}\right)^2 {\rm Tr} (X_l^i X_{-l}^i)
+ {1 \over \gs} \sum_{l \not= 0} \left({2 \pi l \over \beta}\right)^2 {\rm Tr} (\bar{\alpha}_l \alpha_l) \\
& & \!\!\!\! - {1 \over \gssb} \sum_l {2 \pi l \over \beta} {\rm Tr} (X_l^i [A_{00},X_{-l}^i])
+ {1 \over \gssb} \sum_{l \not= 0} {2 \pi l \over \beta} {\rm Tr} (\bar{\alpha}_l [A_{00},\alpha_l]) \\
& & \!\!\!\!  {1 \over 2 \gs \beta} \sum_l {\rm Tr} ([A_{00},X_l^i] [A_{00},X_{-l}^i])
- {1 \over 4 \gs \beta} \sum_{l + m + n + o = 0} \!\!\!\!\!\! {\rm Tr} ([X_l^i,X_m^j][X_n^i,X_o^j]) \,.
\eeas
Next we need to choose an action $S_0$ to use as the basis
for the approximation.\footnote{Out of habit we will keep calling it
the Gaussian approximation, even when $S_0$ is not strictly Gaussian.}
Up to now we have always taken $S_0$ to be quadratic in the
fundamental fields.  But this is not appropriate for gauge theory,
because the eigenvalues of $A_{00}$ are angular variables.  So it is
better to work with the holonomy
\[
U = P e^{i \oint d\tau A_0} = e^{i \sqrt{\beta} A_{00}}\,.
\]
The `Gaussian' action we propose to use is
\[
S_0 = - {N \over \lambda} {\rm Tr} (U + U^\dagger) + \sum_l {1 \over 2 \sigma_l^2}
{\rm Tr} (X_l^i X_{-l}^i) - \sum_{l \not= 0} {1 \over s_l} {\rm Tr} (\bar{\alpha}_l
\alpha_l)\,.
\]
Strictly speaking, we propose to use this action to describe the
relative eigenvalues of the $U(N)$ gauge theory.  That is, $U$ really
takes values in the group $SU(N)/{\Bbb Z}_N$.  But we will largely
ignore this distinction, since it does not affect leading large-$N$
counting.

Although the action $S_0$ is not strictly Gaussian, it is soluble, at least
in the large-$N$ limit: the action for $U$ is the one-plaquette model
studied by Gross and Witten \cite{GrossWitten}.  We briefly recall
some results.  The one-plaquette partition function is
\beas
Z_\scriptlap & = & e^{- \beta F_\scriptlap} = \int dU \, e^{{N \over \lambda} {\rm Tr}
(U + U^\dagger)} \\
\beta F_\scriptlap & = & \left\lbrace
\begin{array}{ll}
N^2\left(-{2 \over \lambda} - \half \log {\lambda \over 2} + {3 \over 4} \right) & \quad \lambda \leq 2 \\
\noalign{\vskip 2mm}
-N^2 / \lambda^2 & \quad \lambda \geq 2
\end{array} \right.
\eeas
with a third-order phase transition at $\lambda = 2$.  Diagonalizing
the holonomy, $U = {\rm diag.}(e^{i \alpha_1},\ldots,e^{i \alpha_N})$,
one has the distribution of eigenvalues
\[
\rho_\scriptlap(\alpha) = \left\lbrace
\begin{array}{ll}
{2 \over \pi \lambda} \cos {\alpha \over 2} \sqrt{{\lambda \over 2} - \sin^2 {\alpha \over 2}}
& \quad \lambda \leq 2 \\
\noalign{\vskip 2mm}
{1 \over 2 \pi} \left(1 + {2 \over \lambda} \cos \alpha\right) & \quad \lambda \geq 2
\end{array} \right.
\]
The expectation value of the Wilson loop is
\[
< {\rm Tr} U>_\scriptlap = \left\lbrace
\begin{array}{ll}
N(1 - \lambda/4) & \quad \lambda \leq 2 \\
\noalign{\vskip 2mm}
N/\lambda & \quad \lambda \geq 2
\end{array} \right.
\]
Finally, the 2-point correlators in $S_0$ are given by
\beas
<A_{00\,AB} A_{00\,CD}>_0 & = & \rho_0^2 \delta_{AD} \delta_{BC} \\
<X^i_{l\,AB} X^j_{m\,CD}>_0 & = & \sigma_l^2 \delta^{ij} \delta_{l+m} \delta_{AD} \delta_{BC} \\
<\bar{\alpha}_{l\,AB} \alpha_{m\,CD}>_0 & = & s_l \delta_{lm} \delta_{AD} \delta_{BC}
\eeas
where
\[
\rho_0^2 = {1 \over \beta N} \int d \alpha \, \alpha^2 \rho_\scriptlap(\alpha) =  \left\lbrace
\begin{array}{ll}
{2 \over \beta N} \left[{\rm li}_2\left(1 - {\lambda \over 2}\right)
+ \left(1 - {2 \over \lambda}\right) \log \left(1 - {\lambda \over 2}\right) - 1 \right]
& \lambda \leq 2 \\
\noalign{\vskip 2mm}
{1 \over \beta N} \left({\pi^2 \over 3} - {4 \over \lambda}\right) & \lambda \geq 2
\end{array} \right.
\]
involves a dilogarithm.

The first two terms in the expansion of the free energy are given by
\bea
\label{bosonic:FreeEnergy}
& & \beta F_0 = \beta F_\scriptlap(\lambda) - {N^2 d \over 2} \sum_l \log \sigma_l^2
+ N^2 \sum_{l \not= 0} \log s_l \\
& & <S - S_0>_0 = {N \over \lambda} <{\rm Tr} (U + U^\dagger)>_\scriptlap + {N^2 d \over 2} \sum_l
\left({1 \over \gs} ({2 \pi l \over \beta})^2 \sigma_l^2 - 1\right) \nonumber \\
& & \qquad + N^2 \sum_{l \not= 0} \left({1 \over \gs} ({2 \pi l \over \beta})^2 s_l + 1\right)
+ {N^3 d \over \gs \beta} \, \rho_0^2 \sum_l \sigma_l^2
+ {N^3 d(d-1) \over 2 \gs \beta} \left(\sum_l \sigma_l^2\right)^2\,. \nonumber
\eea
In writing these expressions we have kept only the leading large-$N$
behavior, coming from planar diagrams.

The next step is to choose a set of gap equations.  For this system we
adopt the operator gap equations discussed in section 2.2.  That is,
we will require that $\beta F_0 + <S - S_0>_0$ be stationary with
respect to variations of the parameters $\lambda$, $\sigma_l^2$,
$s_l$.

Let us first consider the gap equation from varying $\sigma_l^2$,
which reads
\[
{1 \over \sigma_l^2} = {1 \over \gs} \left({2 \pi l \over \beta}\right)^2 + {2 N
\over \gs \beta} \rho_0^2
+ {2 N (d-1) \over \gs \beta} \sum_l \sigma_l^2\,.
\]
The solution is $\sigma_l^2 = \gs / ((2 \pi l / \beta)^2 + m^2)$, parameterized by
an effective thermal mass
\be
\label{eq:YMmass}
m^2 = {2 N \over \beta} \rho_0^2 + 2 (d-1) R_{\rm rms}^2\,.
\ee
Here we have introduced the `size' of the system, defined by the range
of eigenvalues of the matrices $X_i$.
\bea
R_{\rm rms}^2 & = & {1 \over N} {\rm Tr} <(X^i(\tau))^2> \qquad \hbox{\rm (no sum on $i$)}
\nonumber \\
& = & {N \over \beta} \sum_l \sigma_l^2 \nonumber \\
\label{eq:YMRrms2}
& = & {\gs N \over 2 m \tanh (\beta m / 2)}\,.
\eea

Next we consider the gap equation for the ghost propagator, which reads
\[
{1 \over s_l} = - {1 \over \gs} \left({2 \pi l \over \beta}\right)^2 \qquad l \not= 0 \,.
\]
Thus in this approximation the ghosts are free and decoupled.  At leading order they
make a trivial contribution to the partition function, since in $0+1$ dimensions
one has $\det' \partial_\tau = 1$.

Finally, we have the gap equation from varying $\lambda$, which reads
\be
\label{eq:YMlambda} \left\lbrace
\begin{array}{ll}
1 - \left(1 - {2 \over \lambda}\right) \log \left(1 - \lambda \over 2\right)
= \gs N \beta / 4 d R_{\rm rms}^2 & \quad \lambda \leq 2 \\
\noalign{\vskip 2mm}
\lambda = \gs N \beta / 2 d R_{\rm rms}^2 & \quad \lambda \geq 2\,.
\end{array} \right.
\ee

By making use of the gap equations, the expressions (\ref{bosonic:FreeEnergy})
for the free energy can be simplified to
\bea
& & \beta F_0 = \beta F_\scriptlap(\lambda) + N^2 d \log (2 \sinh(\beta m / 2)) \nonumber \\
\label{YM:GaussianBF}
& & <S - S_0>_0 = {2 N \over \lambda} < {\rm Tr} U >_\scriptlap - {\beta N d(d-1) \over 2 \gs}
(R_{\rm rms}^2)^2\,.
\eea

Before studying these equations further, let us note that this
approximation respects 't Hooft large-$N$ counting.  All the factors
of $\gs$ and $N$ which appear in these equations are dictated by 't
Hooft counting plus dimensional analysis, and they can be eliminated
by introducing the following rescaled dimensionless quantities:
\bea
& & \tilde{\beta} = \beta (\gs N)^{1/3} \nonumber \\
& & \tilde{\lambda} = \lambda \nonumber \\
& & \tilde{\rho}_0^2 = N \rho_0^2 / (\gs N)^{1/3} \qquad
    \tilde{\sigma}_l^2 = N \sigma_l^2 / (\gs N)^{1/3} \qquad
    \tilde{s}_l = N s_l / (\gs N)^{1/3}  \nonumber \\
& & \widetilde{R}_{\rm rms}^2 \equiv {1 \over \tilde{\beta}} \sum_l \tilde{\sigma}_l^2
                              = R_{\rm rms}^2 / (\gs N)^{2/3} \nonumber \\
\label{bosonic:scalings}
& & \widetilde{\beta F} = \beta F / N^2\,.
\eea
The 't Hooft coupling $\gs N$ acts as a unit of $({\rm energy})^3$.
The effective dimensionless coupling is then $\gs N / T^3$, so the
system is strongly coupled at low temperatures.

The system of equations (\ref{eq:YMmass}), (\ref{eq:YMRrms2}),
(\ref{eq:YMlambda}) suffices to determine $\lambda$ as a
function of the temperature.  The results for the case of $d = 9$
scalar fields are plotted in Fig.~7.  Note that a Gross-Witten
transition occurs at $\beta = 9.0$, the temperature at which $\lambda$
passes through $2$.  The approximation suggests that the transition is
weakly second-order: the second derivative of $\beta F_0 + <S -
S_0>_0$ jumps by -0.0053 across the transition.\footnote{The fact that
the first derivative is continuous across the transition is guaranteed
by the form of the gap equations that we are using.}  But this may be an artifact
of the approximation; a third order phase transition would seem more
plausible.

\begin{figure}
\epsfig{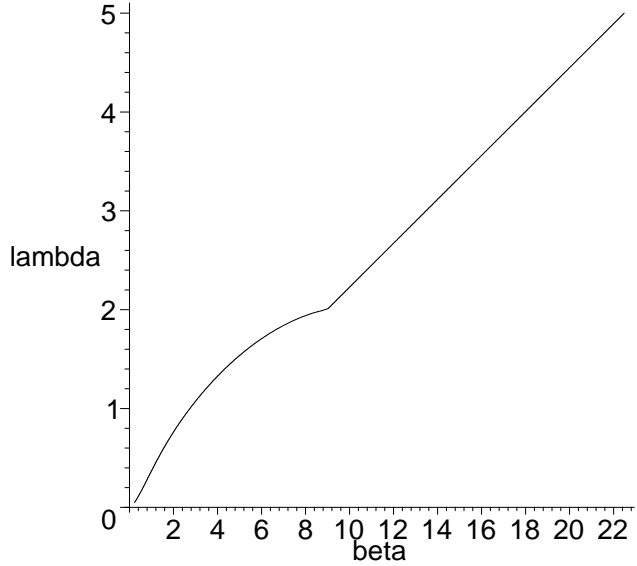}
\caption{$\lambda$ vs.~$\beta$ for bosonic gauge theory with $9$ scalar fields.
The temperature is measured in units of the 't Hooft coupling.}
\end{figure}

At high temperatures ($T > (\gs N)^{1/3}$, or in dimensionless terms
$\tilde{T} > 1$) the Gaussian approximation gives the behavior (with $\beta F \approx
\beta F_0 + <S - S_0>_0$)
\bea
\lambda & \approx & \left({d - 1 \over d(2 d -1)}\right)^{1/2} \tilde{\beta}^{3/2} \nonumber \\
R_{\rm rms}^2 & \approx & (\gs N)^{2/3} \left({2 d - 1 \over 4 d (d-1)}\right)^{1/2}
\tilde{\beta}^{-1/2}
\nonumber \\
\label{bosonic:hightemp}
\beta F & \approx & {3 \over 4} N^2 (d-1) \log \tilde{\beta} \\
E & \approx & {3 \over 4} N^2 (d-1) T \nonumber
\eea
This is the expected result in the high temperature regime: the
$N^2(d-1)$ physical degrees of freedom making up the matrices $X^i$
are excited, with an energy per degree of freedom of order the
temperature.

At low temperatures ($T < (\gs N)^{1/3}$, or in dimensionless terms
$\tilde{T} < 1$) the Gaussian approximation gives
\bea
\lambda & \approx & {(d - 1)^{1/3} \over d} \tilde{\beta} \nonumber \\
R_{\rm rms}^2 & \approx & { (\gs N)^{2/3} \over 2 (d-1)^{1/3}} \nonumber \\
\label{bosonic:lowtemp}
\beta F & \approx & {3 \over 8} N^2 d (d-1)^{1/3} \tilde{\beta} \\
E & \approx & {3 \over 8} N^2 d (d-1)^{1/3} (\gs N)^{1/3} \nonumber
\eea
The physics here is equally simple: the $N^2d$ oscillators have all
been frozen into their ground state.  We see that the ground state
energy of an individual oscillator is very large, of order $(\gs
N)^{1/3}$, while the ground state wavefunction of an individual
oscillator is extremely narrow, being a Gaussian of width
\[
<(\Delta X^i_{AB})^2> \, \sim \, (\gs N)^{1/3} / N\,.
\]
It is only when all of the fluctuations are added together in $R_{\rm
rms}$ that the system as a whole has a very large ground state size,
$R_{\rm rms} \sim (\gs N)^{1/3}$.

Before arguing that the Gaussian approximation really does give
reliable results for this system, let us pause to compare the behavior
of this bosonic system to the behavior expected of 0-brane quantum
mechanics \cite{Susskind,Polchinski}.  The full 0-brane system
corresponds to $d=9$ and differs by including sixteen adjoint
fermions.  In the high temperature regime, we expect that the fermions
can be neglected on general grounds (from the path integral point of
view they are squeezed out by their antiperiodic boundary conditions).
So the results (\ref{bosonic:hightemp}) should be a good approximation
to the behavior of 0-branes at high temperature.  On the other hand,
in the low temperature regime the 0-brane system has a dual
supergravity description \cite{imsy}.  The size of the region in which
supergravity is valid $\sim (\gs N)^{1/3}$ is expected to match the
size of the ground state wavefunction of the supersymmetric gauge
theory.  We see that the scale $(\gs N)^{1/3}$ already appears as the
size of the bosonic ground state wavefunction.  So it seems that
including the fermions should not significantly change the size of the
wavefunction.  The fermions do enter in a crucial way in the
supersymmetric theory, however, because they are necessary for
cancelling the enormous bosonic ground state energy which appears in
(\ref{bosonic:lowtemp}).

We will now argue that the Gaussian approximation provides a good
description of this bosonic gauge theory for all values of the
temperature.  Of course the Gaussian approximation is good in the
high-temperature regime, where the gauge theory is weakly coupled and
the action (\ref{bosonic:action}) is approximately Gaussian.  But it
is also a good approximation at low temperatures.  To establish this,
let us examine the second-order corrections to the ground state energy
of the system.

The second-order correction $-\half <(S - S_0)^2>_{\C,0}$ is a sum of
several terms.  But contributions involving the gauge field propagator
can be neglected at low temperatures, because $\rho_0^2 \rightarrow 0$
as $\beta \rightarrow \infty$ (the zero mode of the gauge field
becomes frozen as the circle decompactifies).  Most of the remaining terms
cancel when the gap equations are imposed, leaving just
\beas
& & - \half <(S - S_0)^2>_{\C,0} \\
& & = - {N^4 d(d-1) \over g_{YM}^4 \beta^2} \sum_{l+m+n+o = 0}
        \sigma_l^2 \sigma_m^2 \sigma_n^2 \sigma_o^2 \\
& & = - {g_{YM}^4 N^4 d(d-1) \beta \over 16 m^5 \sinh^4(\beta m/2)}
\left({3 \over 8} \beta m + \half \sinh (\beta m) + {1 \over 16} \sinh (2 \beta m)\right)
\eeas
(the three-loop integral is easiest to do in position space).
Combining this with previous results gives the first three terms
in the reorganized perturbation series for the free energy, which in
the low temperatures regime reads\footnote{Presumably this model
admits an expansion in $1/d$.  It would be interesting to understand
this expansion, and its relation to the Gaussian approximation.}
\[
\beta F = \half N^2 d (d-1)^{1/3} \tilde{\beta} \left[1 - {1 \over 4} - {1 \over 16(d-1)}
+ \cdots\right]\,.
\]
As claimed, the higher order terms in reorganized perturbation theory
make a small correction to the free energy, even in the
strong-coupling, low temperature regime.  Based on the size of the
second order correction, we can claim to have computed the ground
state energy of this system to within roughly one percent accuracy for
the case of $9$ scalar fields.

\subsection{A supersymmetric large-$N$ scalar model}

Consider a set of $N \times N$ Hermitian scalar multiplets $\Phi_a$,
governed by an action which is a truncation of the 0-brane quantum
mechanics (\ref{D0:action}).
\begin{equation}
S = \frac{1}{\gs} \int dt d^2\theta \, {\rm Tr} \left\{ - \frac{1}{2} D^{\alpha}\Phi_{a}
D_{\alpha}\Phi_{a} - \frac{i}{3} f_{abc} \Phi_{a}[\Phi_{b},\Phi_{c}] \right\}
\end{equation}
Here $a = 1,\ldots,7$ is an index in the ${\bf 7}$ of $G_2$ and $f_{abc}$
is a $G_2$-invariant tensor.  We will subsequently set $\gs = 1$.
For a Gaussian action we take
\begin{equation}
S_0 = \sum_l \frac{1}{2\Delta_{l}^{2}} {\rm Tr} \, \phi^{a}_{l} \phi^{a}_{-l}
+ \sum_l \frac{1}{2\epsilon_{l}^{2}} {\rm Tr} \, f^{a}_{l} f^{a}_{-l}
+ \sum_r \frac{1}{g_{r}} {\rm Tr} \, \bar{\psi}^a_r \psi^a_r \,.
\end{equation}
The one-loop gap equations are evaluated in the large $N$ limit, by
keeping only the planar contributions.  We find that the gap equations
have a familiar form
\bea
\frac{1}{\Delta^{2}_{l}} &=& (\frac{2\pi l}{\beta})^2 +\frac{2A}{\beta}
\sum_{m+n=l}\Delta^{2}_{m}\epsilon^{2}_{n} -\frac{2A}{\beta}\sum_{r-s=l}
h_{r}h_{s} \nonumber \\
\frac{1}{\epsilon^{2}_{l}} &=& 1 +\frac{A}{\beta}\sum_{m+n=l}\Delta^{2}_{m}
\Delta^{2}_{n} \nonumber\\
\label{G2:gap}
\frac{1}{h_r} &=& \frac{2\pi r}{\beta} +\frac{2A}{\beta}\sum_{l+s=r}
\Delta^{2}_{l}h_{s}
\eea
where $h_r =-ig_r$ is real and $A=14N$.

In fact, in the Gaussian approximation, this $G_2$-invariant model is
isomorphic to the $\Phi_1 \Phi_2 \Phi_3$ model studied in section 3.1.
The gap equations (\ref{G2:gap}) can be exactly mapped to those of the
$\Phi_1\Phi_2\Phi_3$ model by defining
\begin{eqnarray}
\beta_{\rm new} & = & A^{1/3} \beta \nonumber \\
\Delta^{2}_{\rm new} & = & A^{2/3} \Delta^2 \nonumber \\
\epsilon^2_{\rm new} & = & \epsilon^2 \nonumber \\
h_{\rm new} & = & A^{1/3} h \,. \nonumber
\end{eqnarray}
Restoring units, note that $A$ is essentially the 't Hooft coupling,
so that $\beta_{\rm new}$ is the physical temperature measured in
units of the 't Hooft coupling.  With this redefinition one finds that
$\beta F_{0}$, $<S-S_{0}>_0$ and $<(S-S_{0})^2>_{\C,0}$ for the $G_2$
model transform into the corresponding quantities for the $\Phi_1 \Phi_2 \Phi_3$
model, up to an overall factor of $7 N^2 / 3$.  Thus all the results
of section 3.1 can be taken over and applied to the $G_2$ model.  In
particular, the free energy falls off as $\beta_{\rm
new}^{-1.1}$ at low temperatures.

\section{Conclusions}

In this paper we've discussed an approximation scheme which can be
used to study strongly-coupled quantum mechanics problems.  We've
shown that the approximation can be formulated in a way which respects
supersymmetry, by introducing gap equations for the auxiliary fields.
We've also shown that it can be applied to large-$N$ theories: indeed
't Hooft scaling is automatic, provided that one keeps just the planar
contribution to the gap equations.

We used the approximation to study a number of toy models of scalar
supersymmetric quantum mechanics, and found that it captures the
correct qualitative behavior of the free energy at strong coupling.
Moreover, it provides a quantitative estimate of certain quantities,
such as the ground state energy or low-lying density of states, which
would be difficult to obtain by any other means.

The approximation has difficulty dealing with gauge symmetry, since
the gap equations tend to violate the Ward identities.  Nonetheless,
we were able to study a large-$N$ gauge theory: the bosonic sector of
0-brane quantum mechanics.  The Gaussian approximation provides a
remarkably accurate description of the ground state of this theory.
The approximation also predicts that a Gross-Witten phase transition
occurs when the effective coupling is of order one.

We argued that such a phase transition should be present in a generic
large-$N$ gauge theory, in particular for the full 0-brane quantum
mechanics.  This means that the perturbative gauge theory regime of
the 0-brane quantum mechanics is separated from the supergravity
regime by a mild phase transition.  It also suggests that the
Horowitz-Polchinski correspondence point is associated with a phase
transition.

Of course our motivation for this project was to develop techniques
for understanding 0-brane quantum mechanics.  We have been able to
incorporate many of the necessary ingredients into the approximation,
including supersymmetry and large-$N$ counting.  But an adequate
analysis of the full 0-brane quantum mechanics requires a more
satisfactory treatment of gauge invariance \cite{KLL}.

\bigskip
\centerline{\bf Acknowledgements}
We are grateful to Stanley Deser, Misha Fogler, Gerry Guralnik, Roman
Jackiw, Anton Kapustin, David Lowe, Samir Mathur, Emil Mottola, Martin
Schaden, Vipul Periwal, Marc Spiegelman, Paul Townsend and Daniel Zwanziger for
valuable discussions.  DK wishes to thank NYU for hospitality while
this work was in progress.  The work of DK is supported by the DOE
under contract DE-FG02-90ER40542 and by the generosity of Martin and
Helen Chooljian.  GL wishes to thank the Aspen Center for Physics for hospitality while
this work was in progress.  The work of GL is supported by the NSF
under grant PHY-98-02484.

\appendix
\section{Supersymmetry in quantum mechanics}

In this appendix we review the superspace construction of ${\cal N} =
2$ theories in $0+1$ dimensions.

With ${\cal N} = 2$ supersymmetry we have an $SO(2)$ R-symmetry, with
spinor indices $\alpha,\beta = 1,2$ and vector indices $i,j = 1,2$.
The $SO(2)_R$ Dirac matrices are real, symmetric, and traceless; for
example one can choose $\gamma^1 = \sigma_1$ and $\gamma^2 = \sigma_3$
where $\sigma_1$ and $\sigma_3$ are Pauli matrices.  Given two spinors
$\psi_\alpha$ and $\chi_\alpha$, besides the invariant $\psi_\alpha
\chi_\alpha$, one can construct a second invariant which we will denote
\[
\psi^\alpha \chi_\alpha = {i \over 2} \epsilon_{\alpha\beta} \psi_\alpha \chi_\beta\,.
\]
${\cal N} = 2$ superspace has coordinates $(t,\theta_\alpha)$ where
$\theta_\alpha$ is real.\footnote{Our convention is that complex
conjugation reverses the order of Grassmann variables.}  Denoting
$\theta^2 = \theta^\alpha \theta_\alpha$, we normalize $\int d^2
\theta \, \theta^2 = 1$.

We introduce the supercharges and supercovariant derivatives
\beas
Q_\alpha & = & {\partial \over \partial \theta_\alpha} + i \theta_\alpha {\partial \over
\partial t} \\
D_\alpha & = & {\partial \over \partial \theta_\alpha} - i \theta_\alpha {\partial \over
\partial t}
\eeas
which obey the algebra
\beas
\lbrace Q_\alpha, Q_\beta \rbrace & = & 2 i \delta_{\alpha\beta} \partial_t \\
\lbrace Q_\alpha, D_\beta \rbrace & = & 0 \\
\lbrace D_\alpha, D_\beta \rbrace & = & - 2 i \delta_{\alpha\beta} \partial_t\,.
\eeas
The simplest representation of supersymmetry is a real scalar superfield
\[
\Phi = \phi + i \psi_\alpha \theta_\alpha + f \theta^2
\]
The general action for a collection of scalar superfields $\Phi_a$ is
\beas
S & = & {1 \over g^2} \int dt d^2 \theta \, - \half D^\alpha \Phi_a D_\alpha \Phi_a - W(\Phi) \\
  & = & {1 \over g^2} \int dt \,\, \half \dot{\phi}_a^2 + i \bar{\psi}_a \dot{\psi}_a + \half f_a^2 -
{\partial W \over \partial \phi_a} f_a + {\partial^2 W \over \partial \phi_a \partial \phi_b} \bar{\psi}_a \psi_b
\eeas
where in the second line we have taken complex combinations $\psi_a = {1
\over \sqrt{2}}\left(\psi_{1a} + i \psi_{2a}\right)$, $\bar{\psi} = {1 \over
\sqrt{2}} \left(\psi_{1a} - i \psi_{2a}\right)$.

In some cases we can extend the $SO(2)$ $R$-symmetry to an $O(2)_R$,
where the extra $\integer_2$ (`$R$-parity') acts as $\theta_\alpha
\rightarrow \gamma^2_{\alpha\beta} \theta_\beta$.  Both the measure
$d^2\theta$ and the kinetic term are odd under $R$-parity.  Thus
$R$-parity is a symmetry if the superpotential is odd in $\Phi$, with
$\Phi \rightarrow - \Phi$ under R-parity.

To describe gauge theories we introduce a real connection on superspace
\beas
\nabla_\alpha & = & D_\alpha + \Gamma_\alpha \\
\nabla_t & = & \partial_t + i \Gamma_t
\eeas
with the gauge transformation $\delta_\Lambda \Gamma_\alpha = i \nabla_\alpha \Lambda$,
$\delta_\Lambda \Gamma_t = \nabla_t \Lambda$.  The corresponding field strengths ${\cal F}_{\alpha\beta}$,
${\cal F}_\alpha$ are real superfields given by
\beas
\lbrace \nabla_\alpha, \nabla_\beta \rbrace & = & - 2 i \delta_{\alpha\beta} \nabla_t + {\cal F}_{\alpha\beta} \\
\lbrace \nabla_t, \nabla_\alpha \rbrace & = & {\cal F}_\alpha
\eeas
We impose the so-called conventional constraint $\delta_{\alpha\beta}
{\cal F}_{\alpha\beta} = 0$, which implies that ${\cal
F}_{\alpha\beta}$ is a vector of $SO(2)_R$:
\[
{\cal F}_{\alpha\beta} = 2 \gamma^i_{\alpha\beta} {\cal F}_i\,.
\]
This fixes $\Gamma_t$ in terms of the spinor connection
$\Gamma_\alpha$, but does not constrain $\Gamma_\alpha$, as the only
independent Bianchi identity
\be
\label{N=2:Bianchi}
\nabla_\alpha {\cal F}_i = - i \gamma^i_{\alpha\beta} {\cal F}_\beta
\ee
is satisfied for any $\Gamma_\alpha$.

We write the expansion of $\Gamma_\alpha$ in `linear' components as
\[
\Gamma_\alpha = \chi_\alpha + A_0 \theta_\alpha + X^i \gamma^i_{\alpha\beta} \theta_\beta
+ d \epsilon_{\alpha\beta} \theta_\beta + 2 \epsilon_{\alpha\beta} \lambda_\beta \theta^2 \,.
\]
The fields $X^i$ are physical scalars, while $\lambda_\alpha$ are
their superpartners, $d$ is an auxiliary boson, $\chi_\alpha$ are
auxiliary fermions, and $A_0$ is the 0+1 dimensional gauge field.
These linear component fields are of primary interest: by using them
to formulate the Gaussian approximation we are effectively making a
Gaussian approximation in superspace for the superfield
$\Gamma_\alpha$, and such an approximation is guaranteed to respect
supersymmetry.  But it is also useful to define the `covariant'
component fields
\beas
X^i \vert_{\rm cov} & \equiv & {\cal F}^i \vert_{\theta = 0} \\
& = & X^i + \half \gamma^i_{\alpha\beta} \chi_\alpha \chi_\beta \big\vert_{\rm lin} \\
\lambda_\alpha \vert_{\rm cov} & \equiv & {\cal F}_\alpha \vert_{\theta = 0} \\
& = & \lambda_\alpha + \half \dot{\chi}_\alpha
+ {i \over 2} [A_0,\chi_\alpha] - {i \over 2} \gamma^i_{\alpha\beta} [X^i,\chi_\beta] + {i \over 2} \epsilon_{\alpha
\beta} [d,\chi_\beta] - {i \over 2} [\chi_\alpha,\chi_\beta \chi_\beta] \big\vert_{\rm lin} \\
A_0 \vert_{\rm cov} & \equiv & \Gamma_t \vert_{\theta=0} \\
& = & A_0 + \half \chi_\alpha \chi_\alpha \big\vert_{\rm lin}\,.
\eeas
These covariant component fields transform in the usual way under
conventional gauge transformations, and are neutral under gauge
transformations which depend on $\theta$.  Written in terms of
covariant component fields the SYM action takes a simple form,
\beas
S_{SYM} & = & {1 \over g^2_{\rm YM}} \int dt d^2\theta \, {\rm Tr} \left\lbrace - {1 \over 4}
\nabla^\alpha {\cal F}_i \nabla_\alpha {\cal F}_i \right\rbrace \\
& = & {1 \over \gs} \int dt \, {\rm Tr} \Bigl\lbrace {1 \over 2} D_0 X^i D_0 X^i + {1 \over 4}
[X^i,X^j] [X^i,X^j] \\
& & \qquad \qquad \qquad + {i \over 2} \lambda_\alpha D_0 \lambda_\alpha
- \half \lambda_\alpha \gamma^i_{\alpha \beta} [X^i,\lambda_\beta]\Bigr\rbrace \bigg\vert_{\rm cov}
\eeas
where $D_0 = \partial_t + i A_0$.  In terms of the linear component fields the action would only take
this form in Wess-Zumino gauge.  We must also introduce a ghost multiplet, for
which we adopt the component expansion
\[
C = \alpha + \beta_\alpha \theta_\alpha + \gamma \theta^2
\]
where $\alpha$ and $\gamma$ are complex Grassmann fields and
$\beta_\alpha$ is a complex boson.  To the SYM action we add the ghost and gauge fixing terms
\[
S_{\rm g.f.} + S_{\rm ghost} = {1 \over g^2_{\rm YM}} \int dt d^2\theta \, {\rm Tr} \left\lbrace 
\half D^\alpha \Gamma_\alpha D^2 D^\beta \Gamma_\beta
+ D^\alpha \bar{C} \nabla_\alpha C \right\rbrace\,.
\]
Note that the gauge-fixed action respects $R$-parity.

\section{Solving gap equations numerically}

The gap equations are an infinite set of coupled non-linear equations
which determine the Fourier modes of the propagators.  In general we
must resort to numerical methods in order to solve them.

An important simplification follows from the fact that quantum
mechanics is UV free.  Thus at large momenta the propagators have
free-field behavior, and at any given value of the temperature we only
need to solve the gap equations to determine a finite number of
Fourier modes.  In the moduli space model of section 3.1 we obtained
good results by solving for modes with $-1.8 \beta < l,r < 1.8 \beta$.

We are left with a finite set of equations, which we solved
numerically using the Newton-Raphson method.  Let us briefly summarize
the method.  Consider a system of equations
\be
F_i(x_1,\ldots,x_n) = 0 \qquad i=1,\ldots,n \,.
\ee
We start by guessing a solution $\vec{x}_{\rm
old}$, and expand $\vec{F}$ to first order around $\vec{x}_{\rm old}$.
Solving the linearized equations gives an improved guess
\begin{equation}
\vec{x}_{\rm new}=\vec{x}_{\rm old} + \delta \vec{x}
\end{equation}
where
\be
\delta \vec{x} = - {\bf J}^{-1} \vec{F}_{\rm old} \qquad \quad
{\bf J}_{ij} = \left. \frac{\partial F_i}{\partial x_j}
\right\vert_{\vec{x}_{\rm old}}
\ee
If the initial guess is sufficiently close to a root, this procedure
can be iterated a few times and will rapidly converge.  We found that
five or six iterations gave excellent results in the moduli space
model.  The Newton-Raphson method does suffer from poor global
convergence, but this can be improved using a procedure known as
backtracking \cite{NumericalRecipes}.

Even with backtracking one needs an initial guess that is sufficiently
close to the actual solution.  One strategy is to start out at
high temperatures.  Then the gap equations are dominated by the zero
modes, and one can obtain an approximate solution analytically.  For
example, consider the moduli space model.  At high temperatures ($\beta
\ll 1$) the gap equations are approximately solved by
\begin{eqnarray}
\sigma_0^2 & = & \sqrt{\beta} \qquad \sigma_{l \not= 0}^2 = (\beta / 2 \pi l)^2 \nonumber \\
\tau_0^2 & = & 1/2 \qquad \tau_{l \not= 0}^2 = 1 \\
h_r & = & \beta / 2 \pi r \nonumber
\end{eqnarray}
This can be used as a starting point for Newton-Raphson.
Then one can solve the system at a sequence of successively lower
temperatures, using the solution at one temperature as the initial
guess for the next temperature.  In the moduli space model we found
that increasing $\beta$ by a factor of $1.2$ at each step worked well.

The gap equations can also be solved to determine the large-momentum
behavior of the propagators at any temperature, up to constants that can be extracted
from the numerical solution for the low-momentum modes.  For example, in the moduli
space model, the propagators at large momenta behave as
\bea
\sigma_l^2 & = & {1 \over (2 \pi l / \beta)^2 + m^2 } \nonumber \\
\tau_l^2 & = & {(2 \pi l / \beta)^2 \over (2 \pi l / \beta)^2 + m^2 } \nonumber \\
h_r & = & {2 \pi r / \beta \over (2 \pi r / \beta)^2 + m^2 }
\eea
where
\[
m^2 = \frac{2}{\beta} \sum_{l=-\infty}^{\infty} \sigma_{l}^{2}\,.
\]
Note that the asymptotic forms of the propagators are related in the
expected way, given by the naive supersymmetry Ward identities.

In fact the high momentum behavior is important for computing
the free energy.  Even in quantum mechanics one has to deal with the
divergent quantity $\sum_l \log \sigma_l^2$.  This sum, which appears
in $\beta F_0$, can be defined by
\begin{equation}
\sum_{-\infty}^\infty \log \big[\sigma_l^2 \big((2 \pi l / \beta)^2 + m^2 \big)\big]
- 2 \log (2\sinh(\beta m/2))\,.
\end{equation}
The modified sum converges, so it can be calculated numerically.

\section{Black hole -- SYM correspondence}

We discuss the connection between ten dimensional Schwarzschild black
holes and finite temperature SYM quantum mechanics.  We relate the two
theories by boosting an eleven dimension black string and
recompactifying it on a smaller circle to keep the entropy the same
\cite{HorowitzMartinec,DMRR}.  Although this transformation is not a
symmetry of the theory, it does preserve some of its properties, and
therefore can be used to identify phenomena on the black hole side
with phenomena in the gauge theory.

\subsection{Boosted black string}
We start with the black string solution in eleven dimensions
\begin{eqnarray}
ds_{11}^{2} & = & -h(r)dt^2 + h^{-1}(r)dr^2 +r^2 d\Omega_{8} +dz^2 \nonumber \\
\label{bhs}
h(r) & = & 1-\frac{r_{0}^{7}}{r^7} \,.
\end{eqnarray}
If the eleventh dimension $z$ is compactified on a circle of radius
$R$ one can reduce this metric to get the ten dimensional metric of a
Schwarzschild black hole in type IIA string theory. The black hole
mass and entropy are given by ($l_{p}$ is the eleven dimensional
Planck length)
\begin{eqnarray}
M & = & \frac{\Omega_{8} R r_{0}^{7}}{l_{p}^{9}}\nonumber\\
\label{ms}
S & = & \frac{\Omega_{8} \pi  R r_{0}^{8}}{2 l_{p}^{9}}
\end{eqnarray}
where $\Omega_{8}=\frac{2\pi^{9/2}}{\Gamma(9/2)}$.

Following \cite{HorowitzMartinec,DMRR} we boost in the $z$ direction
(in the covering space) with boost parameter $\alpha$, and recompactify
the new $z$ direction on a circle with radius
\begin{equation}
R'=\frac{R}{\cosh \alpha}.
\end{equation}
This radius is chosen so that the resulting ten dimensional charged
black hole will have the same entropy (\ref{ms}) as the Schwarzschild black hole.

In string frame the charged black hole solution is
\begin{eqnarray}
ds^2 & = & -f^{-1/2}(r) h(r) dt^2 +f^{1/2}(r)[h^{-1}(r)dr^2 +
 r^2 d\Omega_{8}^{2}]\nonumber\\
e^{2(\phi-\phi_{\infty})} & = & f^{3/2}(r)\nonumber\\
\label{cbh}
A_{0}(r) & = & \frac{r_{0}^{7}\sinh \alpha \cosh \alpha}{r^7 +r_{0}^{7}
 \sinh^{2} \alpha} \\
f(r) & = & 1 + \frac{r_{0}^{7} \sinh^{2} \alpha}{r^7} \nonumber
\end{eqnarray}
We will eventually take the limit $\alpha \rightarrow \infty$ with
$r_{0}$ fixed (this limit was also considered in \cite{hks,er,aeh}), so we
won't distinguish between $\sinh \alpha$ and $\cosh \alpha$.  In this
limit the number of unit charge D0-branes is given by dividing the ADM
mass of the black hole by $1/R'$,
\begin{equation}
N=\frac{7}{8}MR.
\label{nmr}
\end{equation}

Now let's consider the string theory parameters that come from this
compactification. Our conventions are
\be
R = g_{s} l_{s} \qquad \quad
l_{s}^{2} = \frac{l_{p}^{3}}{R} \frac{1}{(2^4 \pi ^7 )^{1/3}}\,.
\ee
One finds (string parameters are labeled by a tilde)
\begin{eqnarray}
\tilde{l}_{s}^{2} & = & \frac{l_{p}^{3}\cosh \alpha}
{R(2^4 \pi^7)^{1/3}}\nonumber\\
\tilde{g}_{s} & = & \left(\frac{R}{l_p \cosh \alpha}\right)^{3/2}
(2^4 \pi^7)^{1/6} \\
\tilde{g}_{YM}^{2} & \equiv & \frac{1}{4\pi^2} \tilde{g}_{s} \tilde{l}_{s}^{-3}=
\frac{(2^4 \pi^7)^{1/3} R^3}{4\pi^2 l_{p}^{6}}\cosh^{-3} \alpha. \nonumber
\end{eqnarray}

The string length grows with $\alpha$, so we rescale all lengths
by $1/\cosh \alpha$. This gives new parameters (labeled by $'$)
\begin{eqnarray}
(l_{s}^{'})^2 & = & \frac{l_{p}^{3}}{R(2^4 \pi^7)^{1/3}}
\cosh^{-1} \alpha \nonumber\\
g^{'}_{s} & = & \tilde{g}_{s}\nonumber\\
\label{p2}
(g_{YM}^{'})^2 & = & \frac{R^3}{l_{p}^{6}}\frac{(2^4 \pi^7)^{2/3}} {4\pi^2} \\
r' & = & r/\cosh \alpha. \nonumber
\end{eqnarray}
Now taking the limit $\alpha \rightarrow \infty$ brings us into the gauge theory
regime with a finite value for the Higgs vev
\begin{equation}
U\equiv\frac{r'}{(l_{s}^{'})^2}=r/l_{s}^{2}.
\label{p3}
\end{equation}
On the supergravity side the change of scale means that 
\begin{equation}
ds^{'2}=\frac{1}{\cosh^{2} \alpha}ds^{2}\,.
\end{equation}
Using equations (\ref{ms},\ref{nmr},\ref{p2},\ref{p3}) we can write
\begin{eqnarray}
& & \frac{r_{0}^{7} \sinh^{2} \alpha}{r^7} =
\frac{d_{0}N g_{ym}^{'2}}{U^7 l_{s}^{'4}} \nonumber\\
& & \frac{r_{0}^{7}}{r^7} =
\frac{9}{16} a_{0} U^{-7}M g_{ym}^{'4}=\frac{U_{0}^{7}}{U^7} \equiv h(U)-1
\label{para1}
\end{eqnarray}
where $d_{0}=\Gamma (7/2) 2^7 \pi^{9/2}$ and 
$a_{0}=\frac{\Gamma(9/2) 2^{11} \pi^{13/2}}{9}$.

As $\alpha \rightarrow \infty$ note that $l_{s}^{'} \rightarrow 0$, so
the metric becomes (defining $t = t^{'}/\cosh \alpha$, $e^2 = g_{YM}^{'2} N$)
\begin{equation}
ds^{'2}=l_{s}^{'2}[-\frac{U^{7/2}}{d^{1/2}_{0} e}h(U)dt^2+h^{-1}(U)
\frac{d^{1/2}_{0}e}{U^{7/2}}dU^2 +
d^{1/2}_{0}eU^{-3/2}d\Omega_{8}^{2}]\,.
\end{equation}
This is indeed the near horizon geometry of a 0-brane black hole.
This analysis is closely related to \cite{seib}.

\subsection{Parameter identification}

To give a precise identification between the Schwarzschild black hole
and the gauge theory we recall that a near extremal black hole has the
following energy density, temperature and entropy density \cite{imsy}.
\begin{eqnarray}
E & = &  a_{p}^{-1} e^{-4} N^2 U_{0}^{7-9}\nonumber\\
eT & = & c_{p}U_{0}^{(5-p)/2} \\
S & = & h_{p}e^{-3} N^2 U_{0}^{(9-p)/2} \nonumber
\label{therobh}
\end{eqnarray}
Here $e^2 = g_{YM}^{2} N$, $U_{0}$ is the position of the horizon in
Poincar\'e coordinates, and $p+1$ is the worldvolume dimension of
the gauge theory.  The constants appearing in these equations are
\begin{eqnarray}
a_{p} & = & \frac{1}{9-p}\Gamma (\frac{9-p}{2}) 2^{11-2p} 
\pi^{\frac{13-3p}{2}}\nonumber\\
h_{p} & = & (7-p)^{-1/2}\Gamma^{-\frac{1}{2}} (\frac{9-p}{2}) 
2^{p-4} \pi^{\frac{3p-13}{4}}\\
c_{p} & = & (7-p)^{3/2}\Gamma^{-\frac{1}{2}} (\frac{9-p}{2}) 
2^{p-6} \pi^{\frac{3p-13}{4}}\,. \nonumber
\end{eqnarray}

The above equations are only reliable if the supergravity
approximation is valid near the horizon of the black hole -- that is,
if the dilaton is small and the curvature is small in string units.
The condition for small curvature near the horizon is $U_{0}^{3-p} \ll
e^{2}$, and the point where the curvature is order one is $T \sim
U_{0}$.  As a passing observation note that while the thermodynamics
depends on $p$, one has a relation $E \sim e^{-2} N^2 T^2 U_{0}^{2}$
which holds independent of $p$, aside from the exact numerical factor.

Now we can make identifications between the Schwarzschild black hole
and the gauge theory.  We consider the case $p=0$.
On the Schwarzschild black hole side there are three parameters
$M$, $R$, $l_{p}$.  On the gauge theory side there is $N$, $E$,
$g^2_{YM}$.  By equating the thermodynamics we obtain the relations
\begin{eqnarray}
& & E = \frac{9}{16}M \nonumber\\
& & g_{YM}^{2} = \frac{(2^4 \pi^7)^{2/3}}{4\pi^2}\frac{R^3}{l_{p}^{6}}\nonumber\\
& & N = \frac{7}{8}RM \nonumber\\
\label{suym}
& & T_{\rm gauge}=T_{\rm Schwarzschild}=\frac{7}{4 \pi r_{0}} \\
& & U_{0}=\frac{(2^4 \pi^7)^{1/3}R r_{0}}{l_{p}^{3}}\nonumber\\
& & S_{\rm gauge}=S_{\rm Schwarzschild} \nonumber
\end{eqnarray}

\subsection{Transition points}

Let's see what the different regimes of the gauge theory correspond to
for the Schwarzschild black hole.  It is convenient to use (we drop
numerical factors)
\begin{eqnarray}
M & = & E = N^2 e^{-6/5}T^{14/5} \nonumber\\
R & = & N^{-1}e^{6/5}T^{-14/5}\nonumber\\
S & = & N^2 e^{-6/5}T^{9/5} \\
l_{p}^{2} & = & N^{-2/3}e^{8/15}T^{-14/5} \nonumber
\end{eqnarray}

There is one issue we would like to comment on.  The specific heat of
a Schwarzschild black hole is negative, while the gauge theory clearly
has a positive specific heat.  Nonetheless their thermodynamics can be
identified, since we hold different quantities fixed when we take
derivatives with respect to the temperature (the temperature is the
same on both sides). In the gauge theory we hold $N$ and $g^2_{YM}$
fixed, but on the supergravity side we hold $R$ and $l_{p}$ fixed.
This is responsible for the difference. Thus the negative specific
heat of the Schwarzschild black hole is not in contradiction with its
description in terms of thermodynamics of an ordinary field theory
(see also \cite{KlebanovSusskind}).

Now let's look at some transition points:
\begin{enumerate}

\item When $R=r_{0}$ the black string becomes unstable and turns into
an eleven dimensional black hole \cite{GL}.  Our
formulas are only appropriate for $R < r_{0}$. At this point one finds
\begin{eqnarray}
S_1 & = & N \nonumber\\
R_1 & = & N^{5/9}e^{-2/3}\nonumber\\
T_1 & = & e^{2/3}N^{-5/9} \\
l_{p}^{2} & = & e^{-4/3}N^{2/9}\,. \nonumber
\end{eqnarray}
This is the BFKS point \cite{BFKS,BFKS2}.

\item At $R=l_p$ the black string is better described as a ten
dimensional black hole and the string theory has $g_s =1$.  At this point
\begin{eqnarray}
S_2 & = & N^{8/7}\nonumber\\
R_2 & = & N^{1/3}e^{-2/3} \\
T_2 & = & e^{2/3}N^{-10/21}\,. \nonumber
\end{eqnarray}

\item Consider the point where the Schwarzschild black hole has a
horizon size of the order the string length,
$r^{2}_{0}=l^{2}_{s}=l_{p}^{3}/R$.  We find the corresponding gauge
theory parameters
\begin{eqnarray}
S_3 & = & N^2\nonumber\\
R_3 & = & e^{-2/3}N^{-1}\nonumber\\
T_3 & = & e^{2/3} \\
g_{s} & = & \frac{1}{N}\,. \nonumber
\end{eqnarray}
This is the point where the curvature at the horizon of the charged
black hole becomes of order one.  Beyond this point the effective
coupling of the gauge theory is small. We have argued in this paper
that at this point there is a phase transition.  We now see that this
transition occurs at the Horowitz-Polchinski correspondence point,
where the Schwarzschild black hole turns into an elementary string
state \cite{HorowitzPolchinski}.
\end{enumerate}


\end{document}